\documentclass[aps,prl,showpacs,twocolumn,superscriptaddress,floatfix]{revtex4-2}
\usepackage[utf8]{inputenc}
\newcommand{\figurescale}{1}
\usepackage{verbatim}
\usepackage{amsmath}
\usepackage{mathtools}

\DeclarePairedDelimiterX\braket[2]{\langle}{\rangle}{#1 \delimsize\vert #2}

\usepackage[pdftex]{graphicx}
\usepackage[separate-uncertainty=true]{siunitx}
\DeclareSIUnit{\rpm}{rpm}

\usepackage[colorlinks=true,urlcolor=blue,linkcolor=blue,citecolor=blue]{hyperref}
\usepackage[usenames,dvipsnames]{xcolor}
\usepackage[T1]{fontenc}
\usepackage{placeins}
\usepackage{nicefrac}
\usepackage{braket}
\usepackage{pdfcomment}
\usepackage{xcolor}
\usepackage{braket}
\usepackage[normalem]{ulem}

\usepackage{lineno}

\begin{document}

%############################## TITLE #########################################
%\title{The Structural, Vibrational, and Electronic Properties of Defects in 2D CrSBr}
% Optional titles
\title{Probing defects and spin-phonon coupling in CrSBr via resonant Raman scattering}
%##############################################################################
%
%############################ AUTHORS #########################################
\author{K.~Torres}\email{kierstin@mit.edu}
\affiliation{Department of Materials Science and Engineering, Massachusetts Institute of Technology, Cambridge, Massachusetts 02139, USA}
\author{A.~Kuc}%\email{a.kuc@hzdr.de}
\affiliation{Helmholtz-Zentrum Dresden-Rossendorf, Abteilung Ressourcen\"okologie, Forschungsstelle Leipzig, Permoserstr. 15, 04318 Leipzig, Germany}
\author{L.~Maschio}%\email{lorenzo.maschio@unito.it}
\affiliation{Dipartimento di Chimica and NIS Centre of Excellence, Universit\`a di Torino, via P. Giuria 5, I-10125 Turin, Italy}
\author{T.~Pham}%\email{thangpt88@gmail.com}
\affiliation{Department of Materials Science and Engineering, Massachusetts Institute of Technology, Cambridge, Massachusetts 02139, USA}
\author{K.~Reidy}%\email{kareidy@mit.edu}
\affiliation{Department of Materials Science and Engineering, Massachusetts Institute of Technology, Cambridge, Massachusetts 02139, USA}

\author{L.~Dekanovsky}%\email{dekanovl@vscht.cz}
\affiliation{Department of Inorganic Chemistry, University of Chemistry and Technology Prague, Technická 5, 166 28 Prague 6, Czech Republic}
\author{Z.~Sofer}%\email{zdenek.sofer@seznam.cz}
\affiliation{Department of Inorganic Chemistry, University of Chemistry and Technology Prague, Technická 5, 166 28 Prague 6, Czech Republic}
\author{F.~M.~Ross}\email{fmross@mit.edu}
\affiliation{Department of Materials Science and Engineering, Massachusetts Institute of Technology, Cambridge, Massachusetts 02139, USA}
\author{J.~Klein}\email{jpklein@mit.edu}
\affiliation{Department of Materials Science and Engineering, Massachusetts Institute of Technology, Cambridge, Massachusetts 02139, USA}
%
%
%##############################################################################
%
\date{\today}
%
%##############################################################################
%									ABSTRACT
%##############################################################################
%
\begin{abstract}
Understanding the stability limitations and defect formation mechanisms in 2D magnets is essential for their utilization in spintronic and memory technologies. Here, we correlate defects in mono- to multilayer CrSBr with their structural, vibrational and magnetic properties. We use resonant Raman scattering to reveal distinct vibrational defect signatures. In pristine CrSBr, we show that bromine atoms mediate vibrational interlayer coupling, allowing for distinguishing between surface and bulk defect modes. We show that environmental exposure causes drastic degradation in monolayers, with the formation of intralayer defects. Through deliberate ion irradiation, we tune the formation of defect modes, which we show are strongly polarized and resonantly enhanced, reflecting the quasi-1D electronic character of CrSBr. Strikingly, we observe pronounced signatures of spin-phonon coupling of the intrinsic phonon modes and the ion beam induced defect modes throughout the magnetic transition temperature. Overall, we demonstrate that CrSBr shows air stability above the monolayer threshold, and provide further insight into the quasi-1D physics present. Moreover, we demonstrate defect engineering of magnetic properties and show that resonant Raman spectroscopy can serve as a direct fingerprint of magnetic phases and defects in CrSBr.
\end{abstract}
%
%##############################################################################
%
\maketitle
%
%###############################################################################
%								MAIN TEXT
%###############################################################################
%

\textbf{Introduction.} Atomically thin van der Waals (vdW) materials are highly sensitive to defects~\cite{Lin2016,Jiang2019} due to their high surface to volume ratio, enabling opportunities to engineer electronic,~\cite{Qiu2013,Lin.2014} optical,~\cite{Klein.2017,Moody.2018,Klein.2019} and magnetic properties through the deliberate introduction of defects.~\cite{Cheng.2013,Fu.2020} While electronic and optical properties of induced defects have been extensively studied in graphene~\cite{Banhart.2010} and semiconducting transition metal dichalcogenides (TMDCs),~\cite{Lin2016, Nan2014, Kang.2014,Amani.2015,Klein.2017,Moody.2018,Klein.2019} tailoring the magnetic properties of such vdW materials through defect engineering remains challenging.~\cite{Cai2015,Guguchia2018,Mathew.2012,Cheng.2013,Fu.2020,Yun.2020,Nguyen.2021,Nisi.2022} Common strategies like introducing defects~\cite{Cai2015, Guguchia2018,Mathew.2012} or magnetic dopants~\cite{Cheng.2013,Fu.2020,Yun.2020,Nguyen.2021,Nisi.2022} are usually accompanied by random disorder that deteriorates desirable correlated electronic and optical properties.~\cite{Fu.2020,Nguyen.2021,Nisi.2022} The ability to engineer magnetic properties is valuable for creating spin textures or magnetic orders for application in nano-spintronics and memory devices.~\cite{Lu.2020, Graham.2020, Beck.2021, Klein.2021} 

Recent discoveries of layered vdW magnets~\cite{Huang.2017, Gong.2017} provide unique means to engineer artificial magnetic phases or textures.~\cite{Lu.2020, Graham.2020, Beck.2021, Klein.2021} However, microscopic studies in archetypal magnets like CrI$_3$~\cite{Huang.2017} and CrGeTe$_3$~\cite{Gong.2017} have been challenging owing to poor air stability.~\cite{Shcherbakov.2018} A particularly promising candidate material is the vdW magnetic semiconductor CrSBr.~\cite{Katscher.1966, Gser.1990, Wang.2019, Telford.2020} CrSBr has created substantial interest for magnetic property engineering due to its anticipated air stability down to the few-layer limit and bulk single-particle gap of $\sim\SI{1.6}{\electronvolt}$ paired with A-type antiferromagnetism.~\cite{Telford.2020,Klein.2022} The combination of magnetic and semiconducting properties manifests in correlated magneto-transport~\cite{Telford.2020, Telford.2022} and magneto-optical effects.~\cite{Lee.2021,Wilson.2021,Klein.2022,Klein.2022a} CrSBr also shows strongly anisotropic electronic transport in multilayers~\cite{Wu.2022} owing to its quasi-1D electronic structure and weak interlayer hybridization.~\cite{Klein.2022} Moreover, a defect-induced magnetic order is observed in magneto-transport~\cite{Telford.2022} and magnetometry measurements.~\cite{Telford.2020,Telford.2022,Boix-Constant.2022,Paz.2022,Klein.2022a} Additionally, narrow optical emission from defects was identified and found to correlate with the defect-induced magnetic order.~\cite{Klein.2022a} These properties, in combination with the recently demonstrated ability to change local structure in CrSBr under the electron beam,~\cite{Klein.2021} provide exciting opportunities for engineering optically accessible magnetic phases.

Critical to ongoing research and applications of CrSBr is a fundamental understanding of defects and material stability limitations down to the monolayer (1L) limit. While in the literature air stability has been suggested,~\cite{Telford.2020, Lee.2021, Klein.2021, Ye2022,Klein.2022,Klein.2022a} some works have observed degradation in few-layer CrSBr.~\cite{Telford.2022, Wu.2022} The exact degradation mechanisms remain unknown and a systematic study on the long term stability of CrSBr from the 1L to the bulk is lacking. Crystal stability is usually critical in halogen containing compounds,~\cite{Huang.2017} with structural, electronic, and magnetic properties affected in device fabrication recipes. Moreover, resonant Raman spectroscopy is commonly used as a direct fingerprint of defects in vdw materials like graphene.~\cite{Malard.2009, Herziger.2014} While some works report on three distinct phonon modes in CrSBr,~\cite{Lee.2021, Klein.2021, Cenker2022, Ye2022, Klein.2022} a joint group theoretical and experimental analysis on the phonon modes in CrSBr is lacking. Moreover, the spin-phonon coupling, layer dependence, and in particular, the effect of defects on the phonon modes have not been addressed in the literature. Such insights are important to further elucidate on the mechanisms of the recently observed exciton-magnon coupling in CrSBr.~\cite{Bae.2022}

%###################### Figure 1 ################################################
\begin{figure*}
	\scalebox{\figurescale}{\includegraphics[width=1\linewidth]{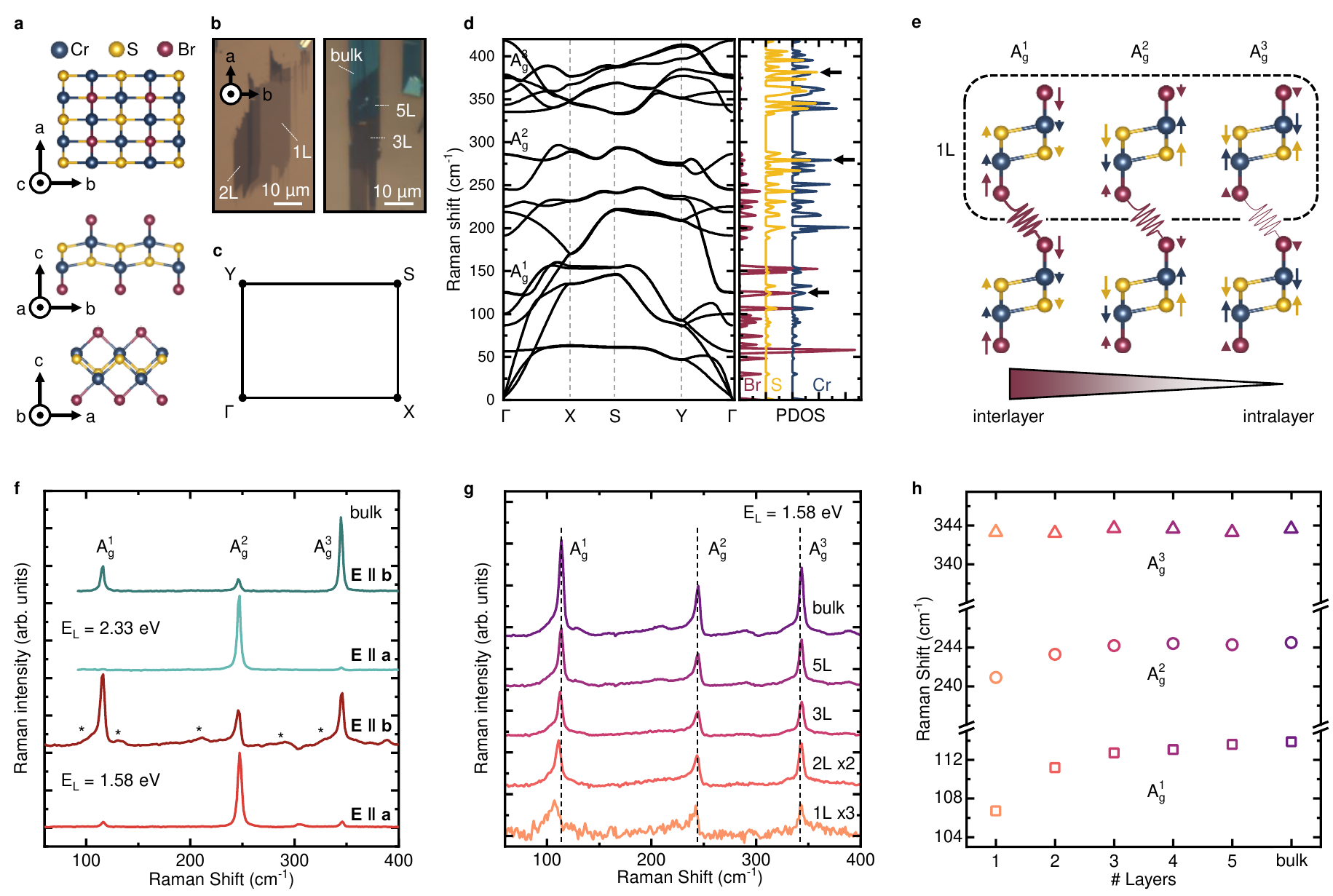}}
	\renewcommand{\figurename}{FIG.|}
	\caption{\label{fig1}
	    \textbf{Phonons and interlayer vibrational coupling in CrSBr.} 
	    \textbf{a}, Schematic crystal structure of CrSBr along the $a$, $b$ and $c$ axes 
	    \textbf{b}, Optical microscope images of 1L to bulk CrSBr exfoliated on SiO$_{2}$/Si substrates.  
     \textbf{c}, First Brillouin zone of 1L CrSBr.
	    \textbf{d}, DFT calculated phonon dispersion of 1L CrSBr and corresponding phonon density of states of chromium (shown in blue), sulfur (yellow) and bromine (red).
	    \textbf{e}, Atomic displacements of the three Raman active A$_{g}$ modes for 1L CrSBr, with two 1L stacked as to display the interlayer coupling. The A$^1_{g}$ and A$^2_{g}$ modes involve bromine atoms that mediate vibrational interlayer coupling, whereas the A$^3_{g}$ mode only shows a slight bromine contribution thus a low interlayer coupling.
	    \textbf{f}, Raman spectra of bulk ($\sim \SI{15}{\nano\meter}$) CrSBr with a non-resonant ($E_L = \SI{2.33}{\electronvolt}$) and resonant ($E_L = \SI{1.58}{\electronvolt}$) excitation. All three $A_{g}$ modes are observed in both excitations with the additional resonant modes, highlighted with *, observed for the resonant excitation.
	    \textbf{g}, Layer dependent Raman spectra from the 1L to the bulk for resonant excitation.
	    \textbf{h}, Corresponding Raman mode positions as a function of layer number.
	    }
\end{figure*}
%##############################################################################

In this work, we uncover the structural and vibrational properties of defects in CrSBr by resonant Raman scattering (RRS) and scanning transmission electron microscopy high-angle annular dark-field (STEM-HAADF) imaging. First, by combining experiment and theory we identify the three dominant out-of-plane phonon modes, A$_g^1$, A$_g^2$, and A$_g^3$, in pristine CrSBr that exhibit a varying degree of interlayer vibrational coupling. Next, we track the long term-stability of CrSBr as well as the impact of defects formed via helium ion (He$^+$) irradiation from 1L to bulk and identify three distinct defect modes, labeled $D1$, $D2$ and $D3$. Mode $D1$ likely correlates with surface related bromine defects from hydrolysis while mode $D2$ and $D3$ likely correlates with intralayer related defects from the chromium/sulfur matrix. We observe a drastic stability difference between 1L and 2L CrSBr explained by observing the $D3$ mode in the 1L. We demonstrate that these defect modes show electronic resonance effects further underpinning the importance of the quasi-1D electronic structure of CrSBr. Finally, in temperature dependent RRS, we observe a pronounced spin-phonon coupling throughout the magnetic transition of CrSBr of both the intrinsic and the He$^+$ beam induced defect modes. Our results demonstrate that 2L and thicker CrSBr is highly air-stable under low humidity and at room temperature. This work provides strategies for controllably inducing defects and magnetic properties in CrSBr, demonstrates RRS as a sensitive probe for detecting defects, dopants, or disorder that breaks crystal symmetry and their related spin-phonon coupling, and underpins CrSBr as an ideal candidate material with potential to study defect engineered magnetic phases.

\textbf{Resonant and non-resonant Raman scattering in CrSBr.} We begin by discussing the vibrational properties of pristine CrSBr, building upon previous work which identified three main phonon modes.~\cite{Lee.2021,Ye2022,Cenker2022} CrSBr possesses an orthorhombic crystal structure with a Pmmn space group and D$_{2h}$ point group. Individual layers are composed of two planes of chromium and sulfur atoms sandwiched by bromine atoms (see Fig.~\ref{fig1}a). CrSBr can readily be exfoliated down to the 1L limit (see Fig.~\ref{fig1}b) and the layer number is determined via atomic force microscopy and optical phase contrast (see SI Fig.~1 and 2, Methods). Flakes display a rectangular, needle structure as a result of the strong crystal anisotropy with the $a$ axis pointing along the direction of the needle. 

As the unit cell of CrSBr contains 6 atoms, we anticipate 18 phonon modes in the first Brillouin zone (see Fig.~\ref{fig1}c). Figure~\ref{fig1}d shows the phonon dispersion and corresponding elementary resolved phonon projected density of states (PDOS) of 1L CrSBr calculated with density functional theory (DFT) (see Methods). The phonon dispersion shows the expected 18 phonon modes. The irreducible representation of phonon modes at the $\Gamma$ point of the Brillouin zone is

%%%
\begin{equation}
    \Gamma = 3 A_{g} \otimes 2 B_{1u} \otimes 3 B_{2g} \otimes 2 B_{2u} \otimes 3 B_{3g} \otimes 2 B_{3u}  ~.
    \label{eq:irr_rep}
\end{equation}
%%%

Only 15 phonon modes are included in the irreducible representation, as the first B$_{1u}$, B$_{2u}$, and B$_{3u}$ modes are acoustic with a zero frequency at the $\Gamma$ point (see Fig.~\ref{fig1}d). Of the remaining modes, the A$_{g}$, B$_{2g}$, and B$_{3g}$ modes are Raman active with the Raman tensors for a Pmmn space group symmetry.~\cite{Loudon.1964}

%%%
\begin{equation}
    \mathcal{R}( A_{g}) = 
     \begin{pmatrix}
a &  &   \\
 & b &   \\
 &  & c  \\
\end{pmatrix}~,~\nonumber
\end{equation}
%%%

%%%
\begin{equation}
    \mathcal{R}( B_{2g}) = 
     \begin{pmatrix}
 &  & e   \\
 & &  &  \\
 e & & \\
\end{pmatrix}
 ~,~\mathcal{R}( B_{3g}) = 
     \begin{pmatrix}
 & & &   \\
 & & f   \\
 & f &   \\ 
\end{pmatrix} ~.~
    \label{eq:R_B}
\end{equation}
%%%

From the Raman tensors, only the three A$_{g}$ modes are anticipated to be detected in first order Raman scattering, also due to the absence of B$_{2g}$ and B$_{3g}$ modes in this particular structure that generally show very weak intensity, further reflected by our calculations (see SI Fig.~4). The representation and Raman active modes agree with that of previously investigated orthorhomic layered magnets, such as TiOX (X = Cl, Br),~\cite{Fausti.2007} FeOCl,~\cite{Bykov.2013} VOCl~\cite{Fausti.2007} and CrOCl.~\cite{Zhang2019}

Figure~\ref{fig1}e depicts the calculated atomic displacements of the A$^1_{g}$, A$^2_{g}$, and A$^3_{g}$ modes in a 2L. The modes correspond to chromium, sulfur, and bromine out-of-plane lattice vibrations but with varying interlayer/intralayer character. This is directly reflected by the elemental contributions to the phonon branches in the PDOS (see Fig.~\ref{fig1}d). Generally branches at lower energies ($< \SI{180}{\per\centi\meter}$) exhibit significantly more contribution from bromine to the PDOS. Since bromine atoms reside at the top and bottom of each individual layer, with sulfur and chromium atoms in the middle (see Fig.~\ref{fig1}a), the vibrational interlayer coupling in multilayer CrSBr is mediated by the bromine atoms with the highest interlayer character for the A$^1_{g}$ mode and the lowest for the A$^3_{g}$ mode (see Fig.~\ref{fig1}e).

Next, we collect Raman spectra using two different laser energies of $E_L = \SI{2.33}{\electronvolt}$ and $E_L = \SI{1.58}{\electronvolt}$ on a bulk flake (see Fig.~\ref{fig1}f). To account for the crystal anisotropy of CrSBr, we co-polarize excitation and detection either along the $a$ axis or the $b$ axis. When exciting at $E_L = \SI{2.33}{\electronvolt}$, the energy is far above the single-particle gap which we refer to as the non-resonant excitation. Here, we observe three modes with clear Lorentzian lineshape that we identify as the A$^1_{g}$, A$^2_{g}$, and A$^3_{g}$ modes with frequencies of $\SI{114}{\per\centi\meter}$, $\SI{244}{\per\centi\meter}$ and $\SI{344}{\per\centi\meter}$.

%###################### Figure 2 ################################################
\begin{figure*}
	\scalebox{\figurescale}{\includegraphics[width=1\linewidth]{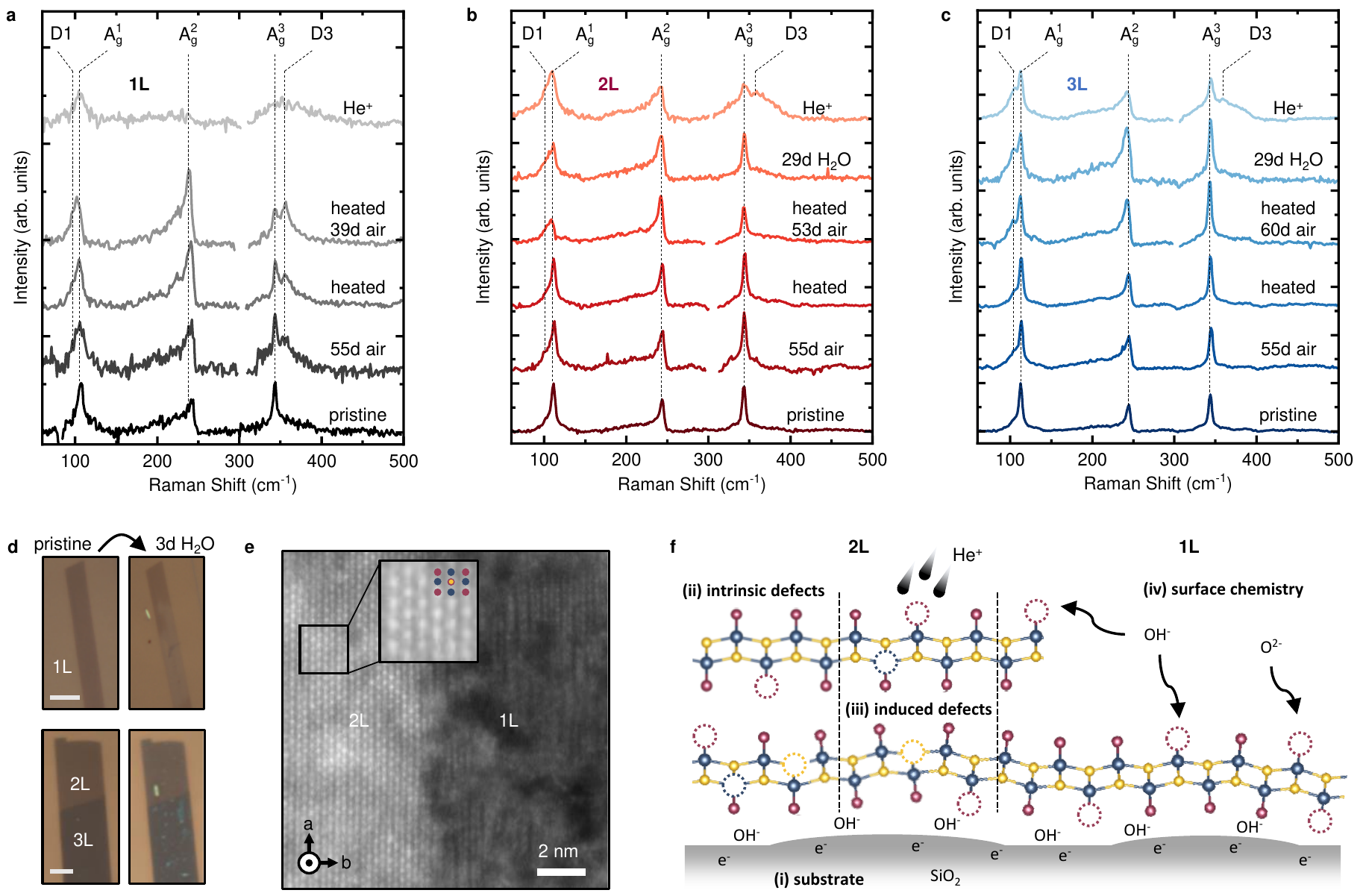}}
	\renewcommand{\figurename}{FIG.|}
	\caption{\label{fig2}
		\textbf{Vibrational and structural signatures of defects in 1L and few-layer CrSBr.}
		Raman spectra of \textbf{a}, 1L, \textbf{b}, 2L, \textbf{c}, 3L pristine CrSBr and after exposure to ambient, water, heating and He$^+$ irradiation. The laser energy is resonant excitation at $E_L = \SI{1.58}{\electronvolt}$. The excitation polarization is linearly polarized with no analyzer in the detection. 
		\textbf{d}, Optical microscope images of 1L, 2L, and 3L CrSBr flakes prior to and after submerging in water for 3 days. Scale bars represent $\SI{5}{\micro\meter}$ for the 1L and $\SI{10}{\micro\meter}$ for the 2L/3L image. 
		\textbf{e}, STEM-HAADF image showing the interface between 1L and 2L CrSBr after heating at $\SI{80}{\celsius}$ for 5 minutes and exposure to water for 1 minute. The 2L shows crystallinity but the 1L strong disorder. The image is collected at a beam energy of $\SI{200}{\kilo\electronvolt}$ and a beam current of $\SI{60}{\pico\ampere}$ and an acquisition time of $\SI{10}{\second}$.
		\textbf{f}, Schematic illustration of the four most prevalent types of defect mechanisms in CrSBr.
		}
\end{figure*}
%##############################################################################

In contrast, for excitation at $\SI{1.58}{\electronvolt}$, the excitation laser is on resonance with the single-particle band gap at $\sim \SI{1.56}{\electronvolt}$~\cite{Klein.2022} and also energetically close to the room temperature exciton emission at $\sim \SI{1.29}{\electronvolt}$ (see SI Fig.~5). We observe that the three A$_{g}$ modes deviate from the pure Lorentzian lineshape. Moreover, additional peaks appear in the spectrum which likely originate from resonant electron-phonon or exciton-phonon coupling effects. The most striking feature is a Fano lineshape in the A$^2_{g}$ mode that has recently been attributed to a van Hove singularity from the quasi-1D electronic conduction band character of CrSBr.~\cite{Klein.2022} For both resonant and non-resonant excitation, we observe that the A$_{g}^{1}$ and A$_{g}^{3}$ modes are polarized with a maximum intensity along the $b$ axis with the A$_{g}^{2}$ mode polarized along the $a$ axis. We also observe that the resonant modes are strongly polarized along the $b$ axis, and vanish for polarization along the $a$ axis (see Fig.~\ref{fig1}f and SI Fig.~6).

The more detailed spectral response for resonant excitation is consistently observed down to the 1L (see Fig.~\ref{fig1}g). Moreover, the A$_{g}^{1}$ and A$_{g}^{2}$ mode frequency are layer dependent with an energy red shift with decreasing layer number (see Fig.~\ref{fig1}h), in contrast to the layer independent energy shift of the A$_{g}^{3}$ mode. Since bromine atoms form the outer layers within each covalently bonded 1L, they also mediate vibrational interlayer coupling between layers. A stiffening with increasing layer number of Raman peaks is commonly observed as a result of increased interlayer vibrational interactions which result in a larger restoring force.~\cite{Lee2010, MolinaSnchez2011} This becomes most apparent from the A$_{g}^{3}$ mode that remains relatively constant suggested by weak bromine lattice vibrations and thereby weak interlayer character (see Fig.~\ref{fig1}e) in excellent agreement with the calculated PDOS (see Fig~\ref{fig1}d).

\textbf{Stability tracked from defect modes.} 
After establishing the structural and vibrational properties of pristine CrSBr, we now investigate the influence of defects on the 1L and few-layer vibrational response for resonant excitation at $E_L = \SI{1.58}{\electronvolt}$. Given that CrSBr may be exposed to air, water, and heating in device fabrication, we collect Raman spectra of samples over several months either exposed to air, heated at $\SI{105}{\celsius}$ for 5 minutes and subsequently exposed to air or submerged in water. We additionally use He$^+$ irradiation with varying concentrations as a more controlled means of forming defects, which has been previously used to engineer defects in 2D materials such as graphene ~\cite{Bell2009,Fox2013} and MoS$_{2}$.~\cite{Fox2015,Klein.2017,Klein.2019,Jadwiszczak2019}

Figure~\ref{fig2}a-c shows the direct comparison of Raman spectra prior to any treatment (pristine), after ambient exposure, heating, heating and subsequent ambient exposure (flake that was initially heated then left in ambient), water exposure, and He$^+$ irradiation. For comparison, spectra are normalized to the A$^1_{g}$ mode. In all spectra, the A$_{g}$ modes remain clearly visible, even after exposure to heat, ambient, and He$^+$ irradiation for all layer numbers. However, several distinct features emerge in the spectra that suggest the presence of defects in CrSBr.

For all spectra, we detect a peak shoulder at $\sim 5-\SI{10}{\per\centi\meter}$ below the $A_{g}^{1}$ mode that we denote as the $D1$ mode. With increasing disorder, this mode increases in intensity and is more prominent in thinner flakes, thereby flakes that exhibit a greater surface to volume ratio. This already suggests that the $D1$ mode is related to surface effects. We note that the $D1$ mode overlaps with the first resonant mode, which is observed most prominently in the bulk spectra. Given that the $D1$ mode is enhanced in thinner flakes and increases in intensity with disorder, this suggests a defect related origin rather than just a resonant mode. However for thicker flakes at low defect levels, differentiating between $D1$ and $R1$ contributions remains challenging.   

Additionally, the 1L after heating and ambient exposure shows a distinct peak, emerging blue shifted from the A$_{g}^{3}$ at $\sim \SI{350}{\per\centi\meter}$ which we label $D3$. This peak is exclusive to the 1L for ambient exposure and heating. Immediately after exfoliation, this peak is not present, but emerges and increases in intensity over time left in air, eventually surpassing the A$_{g}^{3}$ mode in intensity (see SI Fig.~8 a and d). While this mode does not prominently show up in any layer number other than the 1L, He$^+$ irradiation induces a spectrally broader but similar feature in the 2L, 3L, and also bulk CrSBr (see SI Fig.~12). Generally, the data suggest similarities in the defects present in the different layer numbers but also show a strong difference between 1L and the few-layer flakes.

Both air and water exposure yield the $D1$ defect mode, suggesting a similar defect mechanism. Focusing on the peak splitting in the A$_{g}^{1}$ mode, we observe that $D1$ emerges at a more accelerated rate for flakes completely submerged in water compared to those left in air with low humidity $\sim 55\%$ (see SI Fig~11). As water is present in ambient air, it is likely a significant source of surface related defects in CrSBr. This is also suggested by the stronger modification in thinner layers. Moreover, the emergence of the $D3$ mode in the 1L that only shows up in few-layers upon $He^+$ irradiation suggests a different defect origin that is not surface related, which we will discuss in detail below.

To further test the hypothesis of flake degradation in water and the drastic 1L to 2L stability difference, we use high-angle annular dark field scanning transmission electron microscopy (STEM-HAADF) and also take optical microscope images prior to and after water exposure to detect subtle changes in flake contrast. After submerging 1L, 2L, and 3L CrSBr in water for three days, only the 1L shows drastic contrast changes in optical microscope images (see Fig.~\ref{fig2}d). This differs from 2L and 3L flakes, which remain structurally preserved in water with only contamination from the water similar to the SiO$_2$ substrate surface. We underline this observation by collecting STEM-HAADF images of 1L and 2L CrSBr (see Fig.~\ref{fig2}e). During the flake transfer process onto the TEM grid, the CrSBr is heated to $\SI{80}{\celsius}$ for 5 minutes and exposed to water for 1 minute. We observe a very stark difference in stability between the 1L and the 2L. The 2L retains its crystallinity while the 1L becomes more amorphous with visible large defect clusters and holes. 

%###################### Figure 3 ################################################
\begin{figure*}
	\scalebox{\figurescale}{\includegraphics[width=1\linewidth]{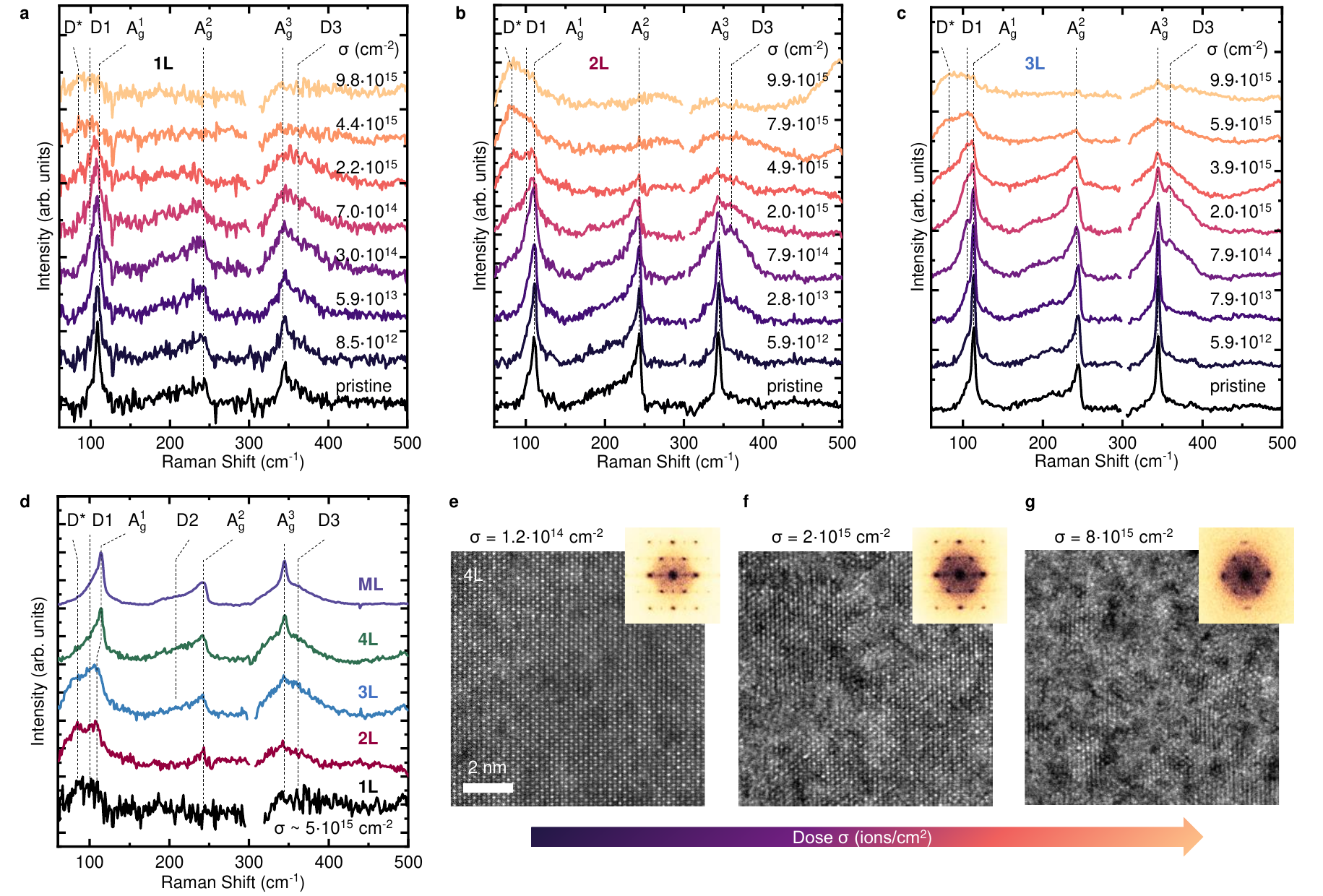}}
	\renewcommand{\figurename}{FIG.|}
	\caption{\label{fig3}
    	\textbf{Defect modes and phonon confinement in He$^+$ irradiated 1L and few-layer CrSBr.} 
    	Raman spectra of \textbf{a}, 1L and \textbf{b}, 2L and \textbf{c}, 3L CrSBr irradiated with different He$^+$ doses and resonant excitation ($E_L = \SI{1.58}{\electronvolt}$). The position of the A$_g$ modes and He$^+$ induced defect modes $D1$, $D3$ and $D^*$ are highlighted, respectively.
    	\textbf{d}, Normalized Raman spectra of 1L to bulk flakes all irradiated at a dose of $\sigma \sim 5 \cdot 10^{15} \SI{}{\per\centi\meter\squared}$. 
    	\textbf{e-g}, STEM-HAADF images of 4L CrSBr irradiated at He$^+$ doses of $\sigma = 2 \cdot 10^{14} \SI{}{\per\centi\meter\squared}$, $1.2 \cdot 10^{15} \SI{}{\per\centi\meter\squared}$ and $8 \cdot 10^{15} \SI{}{\per\centi\meter\squared}$. Inset shows the corresponding FFT.
    	}
\end{figure*}
%##############################################################################

In order to elucidate the defect features in the Raman spectra and understand the differences in stability between 1L and 2L flakes, we now discuss different sources and types of disorder relevant in CrSBr as schematically illustrated in Fig.~\ref{fig2}d: 

(i) Substrate induced effects. The SiO$_{2}$ substrate is known for charge trapping due to dangling bonds at the substrate surface.~\cite{Guo2015, Chae2017} This usually manifests in n-type doping of 2D materials.~\cite{Romero.2008} Moreover, hydroxyl groups on the SiO$_{2}$ from the most common passivation can react with bromine atoms to even form covalent bonds. The surface roughness also generates strain resulting in atomic distortions of the CrSBr flake. Substrate-related effects are expected to be most dominant for the 1L.

(ii) Intrinsic defects from the growth. Our CrSBr crystals exhibit a high intrinsic bromine vacancy concentration of $10^{13} \SI{}{\per\centi\meter\squared}$ and a low concentration of defects tentatively attributed to chromium/sulfur defects of $\sim 3 \cdot 10^{11} \SI{}{\per\centi\meter\squared}$.~\cite{Klein.2022a} From initial experiments, as-grown bulk CrSBr shows n-type behavior in transport studies with a density $> 10^{13} \SI{}{\per\centi\meter\squared}$~\cite{Telford.2022} that is similar to the Br vacancy concentration and is suggestive for the origin of n-type doping in CrSBr.

(iii) Defects within the bulk that are not surface related. Those defects can be induced by heating or ion irradiation. The former is more likely to create bromine vacancies due to the low defect formation energy,~\cite{Klein.2022a} while ion irradiation is expected to substantially increase the concentration of sulfur and chromium vacancy defects, as suggested by Sputtering Range of Ion in Matter (SRIM) simulations (see Methods).~\cite{Ziegler.2010}

(iv) Surface chemistry and defect functionalization. Since the surface of CrSBr consists of bromine atoms, surface chemistry involving water and halogens is a likely mechanism for defect generation. Halogens are commonly known to be reactive especially in presence of water and oxygen that can displace bromine. Hydrolysis from water would result in hydroxyl groups displacing bromine and docking at the vacant bromine sites. Moreover, vacancies in general offer open sites where various atomic or molecular species can be chemisorbed during ambient or water exposure. This likely is similar to the degradation mechanisms in CrI$_{3}$, where hydrolysis and oxidation were demonstrated to result in the formation of Cr(OH)$_{3}$ and CrO$_{3}$ respectively.~\cite{Zhang.2022} Given that in this work, water and air exposure yield a similar defect response, we hypothesize that hydrolysis is the primary defect mechanism observed also given the strong effect when submerged in water. 

For a 1L flake, the entire layer is highly susceptible to surface effects (iv), with one side exposed to the substrate and the other exposed to ambient conditions, whereas for the 2L only half of the layers' surfaces are exposed and for the 3L only one third, etc. Consequently, surface and interface defect mechanisms will have a more pronounced effect on the overall stability and Raman spectra for 1L flakes. Only the 1L displays the $D3$ mode splitting in the A$_{g}^{3}$ mode after air exposure and heating. As the A$_{g}^{3}$ mode corresponds primarily to sulfur and chromium atomic displacements, this suggests that the $D3$ mode similarly arises from intralayer defects related to chromium and sulfur also in line with the observation in 2L and 3L after He$^{+}$ irradiation (see Fig.~\ref{fig2}a-c). It is likely that the large surface to volume ratio in the 1L reduces the barrier to form chromium and sulfur vacancies more readily.

\textbf{Controlled defect generation by He$^+$ irradiation.} The vibrational defect signatures from He$^+$ irradiation suggest that defects related to chromium and sulfur play a role in mode $D3$. We therefore continue to study the impact of He$^+$ irradiation for CrSBr more in depth, since this technique is more likely to substantially increase chromium and sulfur vacancy defects.

We irradiate large areas on 1L to bulk CrSBr with He$^+$ doses ranging from 10$^{12}$ to 10$^{16}$ ions/$\SI{}{\per\centi\meter\squared}$. Figure~\ref{fig3}a-c present Raman spectra at each dose for 1L, 2L and 3L CrSBr (see SI Fig.~12 for 4L and bulk CrSBr). For increasing He$^+$ dose, we observe new features, a consistent peak broadening, and ultimately decreasing Raman signal at critical He$^+$ doses of $\sigma > 10^{16} \SI{}{\per\centi\meter\squared}$. This is expected as higher concentrations of defects reduce the crystallinity of CrSBr. As discussed earlier, bromine point vacancies exist at a high density in 'pristine', non-irradiated bulk CrSBr.~\cite{Klein.2022a} Consequently there is already an imbalance of point vacancy type prior to He$^+$ irradiation. Through He$^+$ irradiation, we expect the formation of more vacancies, both in bulk inner layers (type iii) and at the surface (type iv). According to SRIM simulations,~\cite{Ziegler.2010} the sputter yield for a 1L of CrSBr supported on a SiO$_2$ substrate of chromium, sulfur, and bromine are 0.95\%, 1.25\%, and 1.85\%, respectively after irradiation. Thus, we expect the formation of point vacancies from all atom types. 

% The red shift of the A$_{g}^{1}$ and A$_{g}^{2}$ modes as well as the emergence of the $D1$ defect mode with increasing He$^+$ dose are indicative of a high concentration of bromine vacancies. 

Furthermore, the enhanced intensity of the $D3$ mode with increasing He$^+$ dose is in excellent agreement with chromium/sulfur defects that are expected to scale with the He$^+$ dose. Indeed, we furthermore confirm a decrease in sulfur atom concentration for an increased He$^+$ dose by performing Nano Auger spectroscopy on irradiated multilayer flakes (see SI Fig.~14). In contrast, the change in chromium composition is weak even at high He$^+$ doses. This suggests that chromium is not fully removed but rather displaced into interstitial positions also in agreement with a high defect formation energy.~\cite{Klein.2022a}

Going to high He$^+$ doses ($\sigma > 10^{15}\SI{}{\per\centi\meter\squared}$), we furthermore detect one broad band at $\sim 90-100 \SI{}{\per\centi\meter}$, labeled as $D^*$. While both $D1$ and $D^*$ emerge near the A$_{g}^{1}$ mode, they appear to be distinct defect signatures. This is most prominently observed in the irradiated 2L and 3L where the $D1$ mode can be clearly seen at moderate He$^+$ doses ($\sim 10^{12}-10^{15} \SI{}{\per\centi\meter\squared}$) followed by the broad $D^*$ mode at the highest doses. We systematically observe the same defect modes for all layer thicknesses (see Fig.~\ref{fig3}d) irradiated at a similar He$^+$ dose of $\sim 5 \cdot 10^{15} \SI{}{\per\centi\meter\squared}$. As the multilayer spectra has a higher signal, we observe more clearly an additional defect mode red-shifted from the A$_{g}^{2}$ mode, denoted as $D2$. Due to the overlapping Fano-like lineshape in the A$_{g}^{2}$ mode, this mode is less distinct in thinner flakes yet is likely still present as seen by the broadening near the A$_{g}^{2}$ mode with increasing defect concentration. From our data, it is evident that samples of all thicknesses show a similar defect response but with thinner samples showing higher degradation at the same irradiation dose.

We continue by probing the structural changes of CrSBr collecting STEM-HAADF images of a 4L CrSBr flake irradiated at $\sigma = 2 \cdot 10^{14}$, $1.2 \cdot 10^{15}$, and  $8 \cdot 10^{15} \SI{}{\per\centi\meter\squared}$ (see Fig.~\ref{fig3}e-g). For a low dose ($\sigma = 2 \cdot 10^{14} \SI{}{\per\centi\meter\squared}$) where RRS shows the $D1$ and $D3$ mode but not the $D^*$ mode, we observe point like vacancy defects, visible through intensity changes in individual atomic columns (see Fig.~\ref{fig3}e). In contrast for higher doses ($\sigma > 2 \cdot 10^{15} \SI{}{\per\centi\meter\squared}$) where we observe the $D^*$ mode, the He$^+$ irradiation prompts the formation of larger defect clusters and an increase in disorder. The increasing degree of amorphization is also apparent from the Fast Fourier transform (FFT) of the corresponding STEM images. Here, low dosed flakes yield clear crystal lattice reflections, whereas at higher defect concentrations, reflections blur showing a larger degree of amorphization. Indeed, for an increasing dose the overall crystal structure evolves towards smaller crystalline regions embedded in an amorphous matrix (see Fig.~\ref{fig3}g).

This observation is commonly associated with a phonon confinement with a breakdown of the Raman selection rules. In nanostructured materials where phonons can become confined in real space, the momentum conservation rule is lifted and phonons can spread in momentum space commonly resulting in the activation of additional phonon modes.~\cite{Campbell1986, Richter1981} This intuitive picture is successfully applied to describe defective 2D materials, and it has been used to elucidate the Raman spectra of irradiated MoS$_{2}$~\cite{Mignuzzi2015, Klein.2017, Shi2016} and graphene.~\cite{MartinsFerreira2010} As mentioned earlier, we observe the $D^*$ defect mode upon He$^+$ irradiation in the regime where we observe the formation of nanocrystals in STEM-HAADF (see Fig.~\ref{fig3}g). Therefore, we hypothesize that this mode is either a defect-activated $B_{2g}$ mode at the $\Gamma$ point or due to scattering at another high symmetry point in the Brillouin zone with high PDOS.~\cite{Mignuzzi2015, Klein.2017, Shi2016} 

Given that chromium is magnetic and sulfur is important for mediating exchange interactions in CrSBr,~\cite{Wang.2020} such defects would significantly impact magnetic ordering. We observe that chromium and sulfur defects form upon air exposure for 1L CrSBr, implying less stable magnetic ordering. This is in agreement with a previous observation that the sign and magnitude of the magnetoresistance of a 1L CrSBr device changed after several days of air exposure.~\cite{Telford.2022} Consequently, we anticipate unstable magnetic properties for 1L CrSBr exposed to ambient conditions or heating but stable magnetic ordering for flakes above the 1L threshold. However, our observation of chromium and sulfur related defects at high He$^+$ doses implies that helium ion irradiation can be used to modulate magnetic properties in CrSBr above the 1L threshold. 

The ability to generate localized defects at varying concentrations in CrSBr through helium ion irradiation already shows promise for realizing engineered magnetic phases and textures. Furthermore, we demonstrate that RRS is a powerful tool to detect defect types in varying defect concentration regimes in this material. 

\begin{figure}
	\scalebox{\figurescale}{\includegraphics[width=1\linewidth]{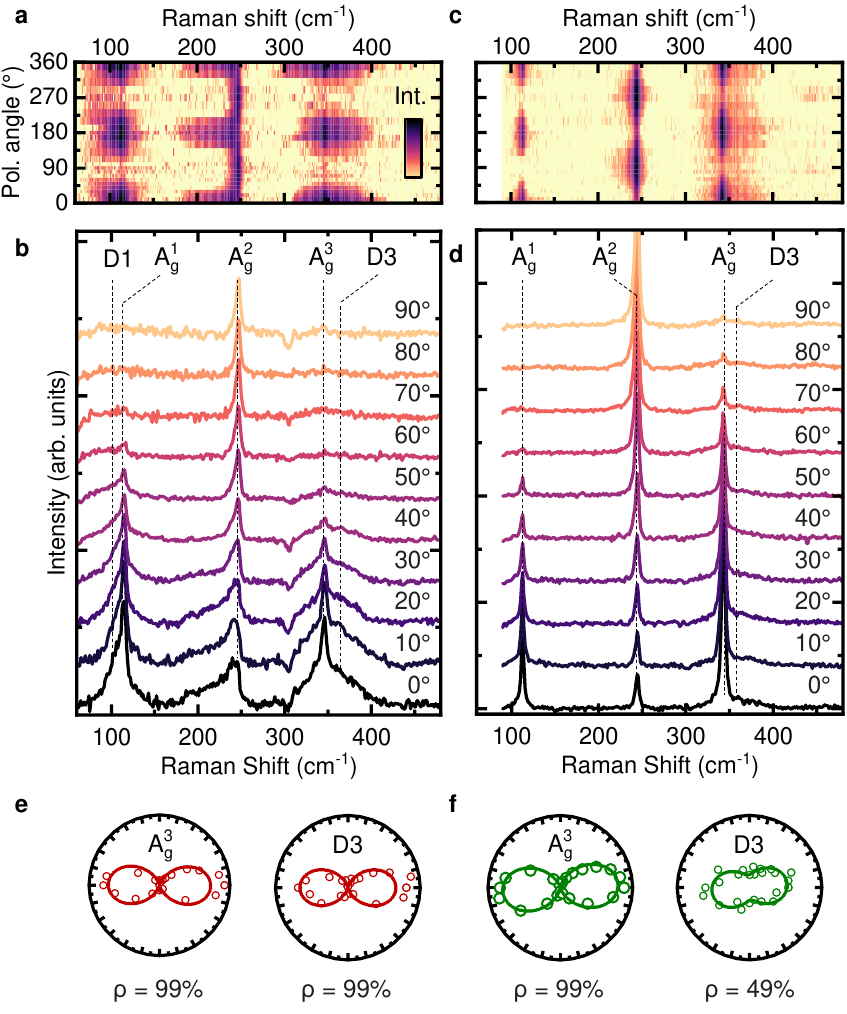}}
	\renewcommand{\figurename}{FIG.|}
	\caption{\label{fig4}
	    \textbf{Resonantly enhanced defect modes in He$^+$ irradiated 4L CrSBr.} 
	    \textbf{a}, False color contour plot of co-linearly polarized Raman of irradiated 4L CrSBr from 0-360 degrees for resonant excitation at $E_L = \SI{1.58}{\electronvolt}$. Angles of 0, 180 and 360 degrees correspond to the $b$ axis while 90 and 270 degree correspond to the $a$ axis. The irradiation dose is $2 \cdot 10^{14}\SI{}{\per\centi\meter\squared}$.
	    \textbf{b}, Selected Raman spectra from 0 to 90 degrees.
	    \textbf{c}, False color contour plot of co-linearly polarized Raman of irradiated 4L CrSBr from 0-360 degrees for non-resonant excitation at $E_L = \SI{2.33}{\electronvolt}$. The irradiation dose is $2 \cdot 10^{14}\SI{}{\per\centi\meter\squared}$.
	    \textbf{d}, Selected Raman spectra from 0 to 90 degrees.
     \textbf{e}, Polar plot of the A$_g^3$ and the $D3$ mode showing polarization along the $b$ axis with a high degree of polarization of $\rho \sim 99\%$, respectively. 
     \textbf{f}, Polar plot for non-resonant excitation shows a polarization along the $b$ axis with a high degree of polarization for the A$_g^3$  mode but a low one for the $D3$ mode ($\rho \sim 49\%$).
		}
\end{figure}

\textbf{Resonant enhancement and polarization of defect modes.} Now that we have established He$^+$ irradiation as technique to deliberately control the defect density, we continue using irradiated CrSBr as a platform to study the effect of the electronic structure onto the resonance physics of the defect modes. In particular, we are now interested in the polarization dependence of the defect modes and the influence of the resonance condition emerging from the quasi-1D electronic structure on the mode intensity. Resonance effects of defect modes are excellent fingerprints in graphene due to the absence of an energy gap.~\cite{Malard.2009, Herziger.2014} Already, we showed that the resonant Raman modes in pristine CrSBr (see Fig.~\ref{fig1}f) are strongly polarized along the $b$ axis (SI Fig. 6). We analogously study vibrational signatures of defects in a He$^+$ irradiated 4L CrSBr at a dose of $\sigma = 2 \cdot 10^{14} \SI{}{\per\centi\meter\squared}$ for resonant and non-resonant excitation where we observe the $D1$ and $D3$ defect modes (in regime of point vacancies). Figure~\ref{fig4}a and b show false color contour plots of polarized Raman spectra ranging from 0\textdegree to 360\textdegree ~taken with resonant and non-resonant excitation. Selected spectra from 0\textdegree ($\parallel$ $b$ axis) to 90\textdegree ($\parallel$ $a$ axis) are depicted in a waterfall representation in Fig.~\ref{fig4}c and d.

Comparing the excitation conditions we make two main observations. First, both the intensity of the $D1$ and $D3$ defect modes are greatly enhanced under resonant excitation, while for non-resonant excitation, the defect peaks are significantly less prominent ($D3$) or even absent ($D1$). Second, both $D1$ and $D3$ defect modes are polarized along the $b$ axis, similar to the $A_g^1$ and $A_g^3$ modes (see Fig.~\ref{fig4}e and f). We can quantify the polarization of the modes by the degree of polarization $\rho = (I_b - I_a)/(I_b + I_a)$ with $I_b$ and $I_a$ as Raman intensity for the laser co-polarized along the $b$ and $a$ axes, respectively. The degree of polarization of the $D3$ mode for resonant excitation is close to unity $\rho \sim 99\%$, while it reaches only $\rho = 49\%$ for non-resonant excitation.

Both intensity enhancement and strong polarization are signatures of resonance effects. Resonance effects are usually rather complex with different potential origins. The strong electronic anisotropy and quasi-1D electronic character of CrSBr around the band edge provide a large number of electronic states that can act as real intermediate electronic transitions for the Raman process.~\cite{Klein.2022} Moreover, the material exhibits excitons with high binding energies in the bulk that show photoluminescence at room temperature (see SI Fig. 5). CrSBr has shown pronounced coupling between excitons and phonons that can also act in exciton-phonon coupling processes.~\cite{Klein.2022} Moreover, the presence of vacancy defects, e.g. V$_{Br}$, V$_S$ or V$_{Cr}$, can induce electronic midgap states with donor and acceptor levels close to band edges.~\cite{Klein.2022a} Such defect levels can provide additional pathways in the scattering process when occupied with an electron or when trapping an exciton.~\cite{Arora.1987,Munz.1980,Berg.1986} Pinning down the exact origin of the enhancement is non-trivial. However, quasi-1D electronic structure, free and localized excitons, and occupied defects can play a role in the resonance effects. The observation of strong polarization along the $b$ axis suggests that the electronic structure of CrSBr dominates the phonon scattering process. Moreover, the clear electronic origin for the resonance effects rules out phonon confinement effects as origin for the $D1$ and $D3$ modes. This is also in excellent agreement with the point vacancy defect regime from STEM-HAADF (see Fig.~\ref{fig3}e).

%###################### Figure 4 ################################################

\begin{figure*}
	\scalebox{\figurescale}{\includegraphics[width=1\linewidth]{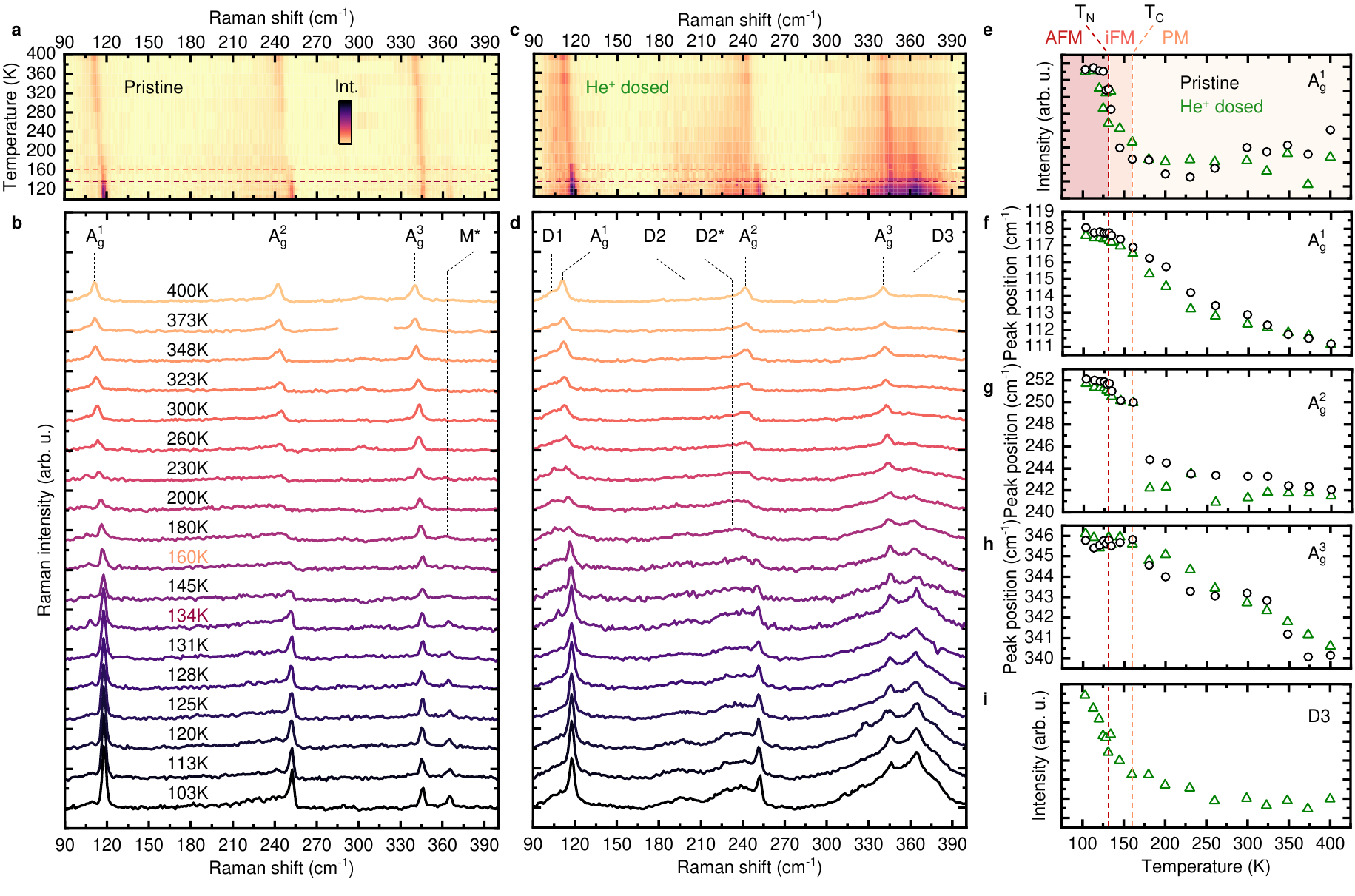}}
	\renewcommand{\figurename}{FIG.|}
	\caption{\label{fig5}
    	\textbf{Pronounced spin-phonon coupling in pristine and He$^+$ irradiated 3L CrSBr.} 
    	\textbf{a}, Contour plot of the Raman temperature evolution of 3L CrSBr collected at $E_L = \SI{1.58}{\electronvolt}$. The dashed horizontal lines mark the Curie ($T_C \sim \SI{160}{\kelvin}$) and N\'eel temperature ($T_N \sim \SI{134}{\kelvin}$).
     \textbf{b}, Corresponding waterfall representation.
     \textbf{c}, Contour plot of the Raman temperature evolution of He$^+$ irradiated 3L CrSBr at a dose of $\sigma = 8 \cdot 10^{14} \SI{}{\per\centi\meter\squared}$. 
     \textbf{d}, Corresponding waterfall representation.
    	\textbf{e}, $A_g^1$ mode intensity of pristine and irradiated material as a function of temperature. The antiferromagnetic (AFM), intermediate ferromagnetic (iFM) and paramagnetic (PM) regime are marked.
     \textbf{f-h}, Temperature evolution of the mode position of $A_g^1$, $A_g^2$ and $A_g^3$ for pristine and He$^+$ irradiated CrSBr.
     \textbf{i}, Temperature dependent intensity of the He$^+$ induced $D3$ defect mode.
    	}
\end{figure*}

\textbf{Spin-phonon coupling in CrSBr.}  After demonstrating the ability to controllably introduce and probe defects in CrSBr, we now study the spin-phonon coupling of both the intrinsic phonon modes and the defect modes via temperature dependent RRS. To this end, we measure a 3L of CrSBr, both pristine and $He^+$ irradiated with a dose of $\sigma = 8 \cdot 10^{14} \SI{}{\per\centi\meter\squared}$. Figures~\ref{fig5}a-d show the full Raman temperature evolution from $\SI{103}{\kelvin}$ to $\SI{400}{\kelvin}$ in a contour plot and the corresponding waterfall representation.

%###################### Figure 5 ################################################
%##############################################################################

Beginning with pristine CrSBr (see Fig.~\ref{fig5}a and b), we observe several distinct changes of the phonon modes with respect to their intensity, lineshape and frequency, in particular throughout the temperature range that is associated with the onset of intralayer correlations (Curie temperature $T_C = \SI{160}{\kelvin}$) and the interlayer correlations (N\'eel transition $T_{N} = \SI{134}{\kelvin}$).~\cite{Telford.2020} Most strikingly, we observe a strong increase of the $A_g^1$ mode intensity below $T_C$ that saturates at $T_N$ (see Fig.~\ref{fig5}e). This is similarly prominent for the $A_g^2$ mode but less pronounced in the $A_g^3$ mode (see Fig.~\ref{fig5}b). Moreover, while all $A$ modes show an expected mode stiffening as the lattice temperature is decreased, all mode frequencies remain constant for $T \leq T_N$. This behavior below $T_N$ deviates from the anharmonic temperature dependence of phonons.~\cite{Granado1999} The $A_g^2$ mode shows the most intricate dependence with a discontinuity of the mode position between $T_C$ and $T_N$ (see Fig.~\ref{fig5}g), a temperature range associated with an intermediate ferromagnetic (iFM) regime.~\cite{Lee.2021} This mode unlike $A_g^1$ and $A_g^3$, shows a maximum intensity along the $a$ axis (see Fig.~\ref{fig4}) and has also been reported to be strongly interacting with the quasi-1D electronic structure of CrSBr.~\cite{Klein.2022} Due to this pronounced electron-phonon interaction, the $A_g^2$ mode likely reacts the strongest to the onset of magnetic correlations that change the electronic band structure when transitioning from the paramagnetic (PM) throughout the iFM phase to the antiferromagnetic (AFM) phase.~\cite{Wilson.2021} Moreover, a new mode appears at $\sim \SI{365}{\per\centi\meter}$ for $T < \SI{230}{\kelvin}$ that becomes more intense for $T < T_C$, labeled as $M^{*}$.

Next, we discuss the irradiated 3L (see Fig.~\ref{fig5}c and d). The $A_g$ modes show the same intensity enhancement and temperature evolution of mode position suggesting that the bulk magnetic properties are maintained for this dose (see Fig.~\ref{fig5}f-h). At room temperature, we observe the expected $D1$, $D2$, and $D3$ defect modes with intensity signatures that are less pronounced as compared to the $A_g$ modes. Cooling below $T_C$, $D2$ and $D3$ show a strong intensity increase (see Fig.~\ref{fig5}i), eventually dominating the spectrum below $T_N$. The defect signature $D2$ that is red-shifted from the $A_g^2$ mode appears as two broad peaks with frequencies of $\sim \SI{195}{\per\centi\meter}$ ($D2$) and $\sim \SI{237}{\per\centi\meter}$ ($D2^*$). The increase contrasts the $D1$ mode which intensity remains mostly unaffected at low-temperature.

The observation of drastic intensity changes below and around the magnetic transition temperatures in combination with the anomalous mode frequencies below $T_N$ suggest that the material exhibits pronounced spin-phonon coupling effects.~\cite{Laverdire.2006,Granado1999,Xu.2022,Du.2019} We attribute the mode enhancement to a spin-dependent contribution to the Raman intensity of the mode that is due to the next-nearest neighbor chromium magnetic correlations that alter the Raman tensor.~\cite{Suzuki.1973} Most importantly, this increase is not limited to the $A_g$ modes but is also observed in the He$^+$ induced defect modes, $D2$ and $D3$, suggesting that the spin-phonon coupling effects are also present for the defect modes. The absence of this enhancement in the $D1$ mode might be due to the predominant Br contribution to the atomic displacement (see Fig.~\ref{fig1}d and e), in contrast to chromium and sulfur for the $A^2_g$ and $A^3_g$ modes. This would suggest an elemental specific dependence of magneto-Raman active defect signatures. The continuous increase of the $D2$ and $D3$ modes below $T_N$ also suggest that the defects experience more of the local spin structure, for example a reduction in local spin fluctuations~\cite{Paz.2022} that can potentially change their mode intensity. The observation of such a strong magnetic correlation of modes and particularly the defect modes is interesting for engineering magnetic phases and for accessing their properties using optical probes like Raman spectroscopy. We further note that we observe similar defect modes as well as the $M^{*}$ mode also in 2L CrSBr at $\SI{113}{\kelvin}$ (see SI Fig. 15) and rule out a symmetry effect as origin from an alternating even/odd layer number. Moreover, we note that the $M^{*}$ mode frequency is similar to $D3$ suggesting that this mode could be from a residual defect concentration in the pristine material, however, we cannot determine the exact origin of this mode at this point.

\textbf{Conclusions and Outlook.} In conclusion, we correlated the vibrational and magnetic response with atomic-level disorder and defects in CrSBr through resonant Raman scattering. We expose CrSBr to a variety of environments revealing rich signatures of defects and their dependence on formation conditions. We show that the most relevant defect mechanism in CrSBr is related to bromine atoms on the surface that are susceptible to displacement via water (hydrolysis), resulting in the $D1$ mode. The observation of the $D3$ mode after air, heating, and water exposure only in the 1L implies that chromium and sulfur defects form more readily at the 1L level. However, above the 1L threshold, CrSBr is highly stable in ambient conditions when exposure to water and heating is limited.

Furthermore, we observe that resonant excitation greatly enhances defect Raman signatures with a strong polarization along the $b$ axis, reflecting the quasi-1D electronic structure of CrSBr. The exact origin of the enhancement is non-trivial but can result from both electron-phonon or exciton-phonon coupling effects. Using a tunable narrow bandwidth laser can help to selectively excite excitonic transitions or other critical points in the rich electronic structure~\cite{Klein.2022} to further understand the distinct resonance effects in this material.

Lastly, we observe a strong coupling between the spin and lattice degrees of freedom that manifests in an anomalous mode dependence throughout the onset of magnetic correlations associated with an intermediate ferromagnetic phase and the antiferromagnetic phase below the N\'eel transition. The spin-phonon coupling effects also extend to deliberately induced defect modes. The rich vibrational properties, material stability, resonant enhancement and magneto-correlated properties open new pathways to engineer magnetic properties in CrSBr by introducing dopants and stoichiometric changes and to use RRS as an optical fingerprint to probe the magnetic phase diagram. 

%
%##############################################################################
%               Acknowledgements & Contributions
%##############################################################################
%
\section{Methods}

\subsection{Crystal synthesis}

We grew CrSBr crystals by direct reaction from chromium (99.99\%, -60 mesh, Chemsavers, USA), bromine (99.9999\%, Sigma-Aldrich, Czech Republic) and sulfur (99.9999\%, Stanford Materials, USA). We mixed the crystals in stochiometric ratio in a quartz ampoule (35x220mm) using 15g of CrSBr and a bromine excess of 0.5g to enhance vapor transport. We pre-reacted the material in an ampoule using a crucible furnace at 700 °C for 12 hours, while we kept the second end of the ampoule below 250 °C. We repeated the heating procedure twice until the liquid bromine disappeared. We then placed the ampoule in a horizontal two zone furnace for the crystal growth. First we heated the growth zone to 900 °C, while we heated the source zone to 700 °C for 25 hours. For the growth, we reversed the thermal gradient and we heated the source zone from 900 °C to 940 °C and the growth zone from 850 °C to 800°C over a period of 7 days. We removed crystals with dimensions up to 5x20 mm from the ampule in an Ar glovebox. 

\subsection{Sample fabrication}

We exfoliated bulk CrSBr flakes using the Scotch tape method onto SiO$_2$/Si chips with an oxide thickness of $\SI{90}{\nano\meter}$. Individual flakes and their thickness are determined by atomic force microscopy and optical phase contrast. We transferred selected flakes to TEM compatible sample grids using cellulose acetate butyrate (CAB) as polymer handle.~\cite{Schneider.2010} After the transfer, we dissolved the CAB in acetone and rinsed the grid in isopropanol before critical point drying.

\subsection{Raman spectroscopy}

We collected the Raman data on a Raman inVia confocal microscope (Renishaw). For our measurements we used two linearly polarized laser lines at $\SI{532}{\nano\meter}$ ($E_L = \SI{2.33}{\electronvolt}$) and $\SI{785}{\nano\meter}$ ($E_L = \SI{1.58}{\electronvolt}$). The magnification of the microscope objective was either 50x or 100x with a laser spot diameter of $\sim \SI{2}{\micro\meter}$ or $\sim \SI{1}{\micro\meter}$. With the exception of the angle resolved polarization measurements, all spectra were acquired with the excitation laser aligned to the $b$ axis. For the polarization resolved measurements we co-aligned excitation and detection polarization. We changed the polarization angle by rotating the sample on a rotation stage. For the measurement we used $\SI{10}{\second}$ integration time and average a total of $20$ spectra to increase the signal to noise ratio. We subtracted a background spectrum taken next to the CrSBr on the SiO$_2$/Si substrate at the same conditions. For the temperature dependent measurements, we used a Linkam THMS600 stage.

\subsection{Helium ion microscopy}

We performed controlled He$^+$ irradiation in an ORION NanoFab (Zeiss). We used the same irradiation approach for both free standing and supported samples. The beam energy was $\SI{30}{\kilo\electronvolt}$, the beam current was $\SI{1}{\pico\ampere}$ with a pixel spacing of $\SI{5}{\nano\meter}$ or $\SI{10}{\nano\meter}$. The dwell time was set to $\SI{1}{\micro\second}$ and for adjusting the ion dose, we varied the number of repeats writing a defined two-dimensional area. In order to ion irradiate the entire region illuminated by the focused laser spot with a diameter of $\sim 1-\SI{2}{\micro\meter}$, for our optical experiments we irradiated a large area of a minimum of $\SI{3}{\micro\meter}$ x $\SI{3}{\micro\meter}$. 

\subsection{STEM imaging}

We performed STEM imaging in a probe-corrected Thermo Fisher Scientific Themis Z G3 $60$-$\SI{200}{\kilo\volt}$ S/TEM operated at $\SI{200}{\kilo\volt}$ with a probe convergence semi-angle of 19 mrad. The probe size of the aberration-corrected electron beam is sub-Angstrom. For STEM-HAADF image acquisition, we used a collection semi-angle of 63-200 mrad. We used a typical beam current of $\SI{60}{\pico\ampere}$ at $\SI{200}{\kilo\volt}$. Data were acquired using Velox software (Thermo Fisher) with a typical frame size of 1024x1024 pixel and a dwell time of $\SI{500}{\nano\second}$/pixel.

\subsection{Nano-Auger electron spectroscopy}

We performed nano-Auger electron spectroscopy (nano-AES) using the PHI Model 700 Nanoprobe (Physical Electronics). We He$^+$ irradiated the samples directly before transfer into the nano-Auger system and pumped to a high vacuum of $10^{-12} \SI{}{\bar}$. Samples are located with the SEM and we collected spectra at selected positions in addition to elemental mappings for Cr, S, Br, O and C.

\subsection{DFT calculations}

We studied bulk CrSBr using the Pmmn space group ($\#$59). We cut out layered models from bulk, keeping the remaining symmetries. We kept all systems at the experimental lattice vectors of bulk CrSBr ($a$ = 3.5059 \AA, $b$ = 4.7703\AA, $c$ = 7.9620\AA;~\cite{Klein.2021} we only optimized atomic positions) employing density functional theory simulations as implemented in {\sc Crystal17} program.~\cite{crystal17} We used the PBE0 hybrid functional~\cite{PBE0} with D3 dispersion correction~{D3} and Gaussian-type POB triple-zeta basis sets~\cite{pob_basis} for all simulations. We investigated all systems in two different magnetic configurations, ferro- and antiferromagnetic, however, we found that vibrational analysis (see below) is not affected by the spin state.
For optimization and vibrational simulations of layers, we used $12\times12$ k-mesh, while for bulk we used a $12\times12\times6$ k-mesh following the Monkhorst-Pack scheme. We simulated phonon dispersion relation and density of states of 1L CrSBr using a finite displacement method on a $3\times3$ supercell. We calculated Raman intensities in a non-resonant regime using a linear response approach.~\cite{maschio2013ab}

\subsection{SRIM simulation}

We simulated the sputter yield for Cr, S and Br atoms for He$^+$ irradiation with an energy of $\SI{30}{\kilo\electronvolt}$ using the Sputtering Range of Ion in Matter (SRIM) software package.~\cite{Ziegler.2010} We simulated the sputter yield for monolayer CrSBr on $\SI{90}{\nano\meter}$ of SiO$_2$ and thick Si ($>\SI{300}{\nano\meter}$ of SiO$_2$). We obtained a sputter yield of 0.95\%, 1.25\%, and 1.85\% for chromium, sulfur, and bromine, respectively. Moreover, we found that the sputter yield did not change for CrSBr with up to a thickness of 10 layers (0.94\%, 1.36\%, and 2.22\%).

\section{Acknowledgements}
J.K. acknowledges support by the Alexander von Humboldt foundation. T.P. and F.M.R. acknowledge funding from the U.S. Department of Energy, Office of Basic Energy Sciences, Division of Materials Sciences and Engineering under Award DE‐SC0019336 for STEM characterization. Z.S. was supported by project LTAUSA19034 from Ministry of Education Youth and Sports (MEYS) and by ERC-CZ program (project LL2101) from Ministry of Education Youth and Sports (MEYS). L.D. was supported by specific university research (MSMT No. 20-SVV/2022). A.K. acknowledges financial supported by the Deutsche Forschungsgemeinschaft (project CRC1415, number 417590517), association with priority program (project SPP2244 (2DMP)), and thanks the high-performance computing center of ZIH Dresden for computational resources. K.R. acknowledges funding and support from a MIT MathWorks Engineering Fellowship and ExxonMobil Research and Engineering Company through the MIT Energy Initiative. We acknowledge fruitful discussions with Matthias Florian and Ursula Wurstbauer. We thank Joachim Dahl Thomsen for assistance with flake transfers. We thank Elisabeth Shaw for assistance with Nano Auger measurements and James M. Daley and Ilya Charaev for continuous support with the helium ion microscope.

\section{Author contributions}
K.T., F.M.R. and J.K. conceived the project and designed the experiments, K.T. and J.K. prepared the samples, K.T. and J.K. performed Raman measurements, K.T. performed Nano Auger experiments, T.P. collected STEM data, J.K. and K.R. performed helium ion irradiation, A.K. and L.M. carried out ab initio simulations, Z.S. and L.D. synthesized CrSBr crystals, K.T. and J.K. analyzed the experimental data, K.T. and J.K. wrote the manuscript with input from all co-authors.

\section{Additional Information}

\subsection{Supporting Information} Determination of flake thickness; Calculated phonon modes; Calculated layer-dependent Raman spectra; Laser excitation energy dependence and resonance condition; Polarization of resonant Raman modes; Raman spectrum background subtraction; Time evolution and stability; Raman spectra of helium ion irradiated 4L and bulk; Nano Auger electron spectroscopy; Low- and room temperature Raman spectra of 2L and 3L.

\subsection{Data availability} The data that support the findings of this study are available from the corresponding author on reasonable request.

\subsection{Code availability} The codes used for data analysis as well as \emph{ab initio} calculations are available from the corresponding author on reasonable request.

\subsection{Competing financial interests} The authors declare no competing financial interests.

\bibliographystyle{naturemag}
\bibliography{full}% Produces the bibliography via BibTeX.

\begin{thebibliography}{10}
\expandafter\ifx\csname url\endcsname\relax
  \def\url#1{\texttt{#1}}\fi
\expandafter\ifx\csname urlprefix\endcsname\relax\def\urlprefix{URL }\fi
\providecommand{\bibinfo}[2]{#2}
\providecommand{\eprint}[2][]{\url{#2}}

\bibitem{Lin2016}
\bibinfo{author}{Lin, Z.} \emph{et~al.}
\newblock \bibinfo{title}{Defect engineering of two-dimensional transition
  metal dichalcogenides}.
\newblock \emph{\bibinfo{journal}{2D Materials}} \textbf{\bibinfo{volume}{3}},
  \bibinfo{pages}{022002} (\bibinfo{year}{2016}).
\newblock \urlprefix\url{https://doi.org/10.1088/2053-1583/3/2/022002}.

\bibitem{Jiang2019}
\bibinfo{author}{Jiang, J.}, \bibinfo{author}{Xu, T.}, \bibinfo{author}{Lu,
  J.}, \bibinfo{author}{Sun, L.} \& \bibinfo{author}{Ni, Z.}
\newblock \bibinfo{title}{Defect engineering in {2D} materials: Precise
  manipulation and improved functionalities}.
\newblock \emph{\bibinfo{journal}{Research}} \textbf{\bibinfo{volume}{2019}},
  \bibinfo{pages}{1--14} (\bibinfo{year}{2019}).
\newblock \urlprefix\url{https://doi.org/10.34133/2019/4641739}.

\bibitem{Qiu2013}
\bibinfo{author}{Qiu, H.} \emph{et~al.}
\newblock \bibinfo{title}{Hopping transport through defect-induced localized
  states in molybdenum disulphide}.
\newblock \emph{\bibinfo{journal}{Nature Communications}}
  \textbf{\bibinfo{volume}{4}} (\bibinfo{year}{2013}).
\newblock \urlprefix\url{https://doi.org/10.1038/ncomms3642}.

\bibitem{Lin.2014}
\bibinfo{author}{Lin, Y.-C.}, \bibinfo{author}{Dumcenco, D.~O.},
  \bibinfo{author}{Huang, Y.-S.} \& \bibinfo{author}{Suenaga, K.}
\newblock \bibinfo{title}{Atomic mechanism of the semiconducting-to-metallic
  phase transition in single-layered {MoS}$_2$}.
\newblock \emph{\bibinfo{journal}{Nature Nanotechnology}}
  \textbf{\bibinfo{volume}{9}}, \bibinfo{pages}{391--396}
  (\bibinfo{year}{2014}).
\newblock \urlprefix\url{https://doi.org/10.1038/nnano.2014.64}.

\bibitem{Klein.2017}
\bibinfo{author}{Klein, J.} \emph{et~al.}
\newblock \bibinfo{title}{Robust valley polarization of helium ion modified
  atomically thin {MoS}$_2$}.
\newblock \emph{\bibinfo{journal}{2D Materials}} \textbf{\bibinfo{volume}{5}},
  \bibinfo{pages}{011007} (\bibinfo{year}{2017}).
\newblock \urlprefix\url{https://doi.org/10.1088/2053-1583/aa9642}.

\bibitem{Moody.2018}
\bibinfo{author}{Moody, G.} \emph{et~al.}
\newblock \bibinfo{title}{Microsecond valley lifetime of defect-bound excitons
  in monolayer {WSe}$_2$}.
\newblock \emph{\bibinfo{journal}{Physical Review Letters}}
  \textbf{\bibinfo{volume}{121}} (\bibinfo{year}{2018}).
\newblock \urlprefix\url{https://doi.org/10.1103/physrevlett.121.057403}.

\bibitem{Klein.2019}
\bibinfo{author}{Klein, J.} \emph{et~al.}
\newblock \bibinfo{title}{Site-selectively generated photon emitters in
  monolayer {MoS}$_2$ via local helium ion irradiation}.
\newblock \emph{\bibinfo{journal}{Nature Communications}}
  \textbf{\bibinfo{volume}{10}} (\bibinfo{year}{2019}).
\newblock \urlprefix\url{https://doi.org/10.1038/s41467-019-10632-z}.

\bibitem{Cheng.2013}
\bibinfo{author}{Cheng, Y.~C.}, \bibinfo{author}{Zhu, Z.~Y.},
  \bibinfo{author}{Mi, W.~B.}, \bibinfo{author}{Guo, Z.~B.} \&
  \bibinfo{author}{Schwingenschlögl, U.}
\newblock \bibinfo{title}{Prediction of two-dimensional diluted magnetic
  semiconductors: Doped monolayer {MoS}$_2$}.
\newblock \emph{\bibinfo{journal}{Physical Review B}}
  \textbf{\bibinfo{volume}{87}} (\bibinfo{year}{2013}).
\newblock \urlprefix\url{https://doi.org/10.1103/physrevb.87.100401}.

\bibitem{Fu.2020}
\bibinfo{author}{Fu, S.} \emph{et~al.}
\newblock \bibinfo{title}{Enabling room temperature ferromagnetism in monolayer
  {MoS}$_2$ via in situ iron-doping}.
\newblock \emph{\bibinfo{journal}{Nature Communications}}
  \textbf{\bibinfo{volume}{11}} (\bibinfo{year}{2020}).
\newblock \urlprefix\url{https://doi.org/10.1038/s41467-020-15877-7}.

\bibitem{Banhart.2010}
\bibinfo{author}{Banhart, F.}, \bibinfo{author}{Kotakoski, J.} \&
  \bibinfo{author}{Krasheninnikov, A.~V.}
\newblock \bibinfo{title}{Structural defects in graphene}.
\newblock \emph{\bibinfo{journal}{{ACS} Nano}} \textbf{\bibinfo{volume}{5}},
  \bibinfo{pages}{26--41} (\bibinfo{year}{2010}).
\newblock \urlprefix\url{https://doi.org/10.1021/nn102598m}.

\bibitem{Nan2014}
\bibinfo{author}{Nan, H.} \emph{et~al.}
\newblock \bibinfo{title}{Strong photoluminescence enhancement of mos$_{2}$
  through defect engineering and oxygen bonding}.
\newblock \emph{\bibinfo{journal}{{ACS} Nano}} \textbf{\bibinfo{volume}{8}},
  \bibinfo{pages}{5738--5745} (\bibinfo{year}{2014}).
\newblock \urlprefix\url{https://doi.org/10.1021/nn500532f}.

\bibitem{Kang.2014}
\bibinfo{author}{Kang, N.}, \bibinfo{author}{Paudel, H.~P.},
  \bibinfo{author}{Leuenberger, M.~N.}, \bibinfo{author}{Tetard, L.} \&
  \bibinfo{author}{Khondaker, S.~I.}
\newblock \bibinfo{title}{Photoluminescence quenching in single-layer {MoS}$_2$
  via oxygen plasma treatment}.
\newblock \emph{\bibinfo{journal}{The Journal of Physical Chemistry C}}
  \textbf{\bibinfo{volume}{118}}, \bibinfo{pages}{21258--21263}
  (\bibinfo{year}{2014}).
\newblock \urlprefix\url{https://doi.org/10.1021/jp506964m}.

\bibitem{Amani.2015}
\bibinfo{author}{Amani, M.} \emph{et~al.}
\newblock \bibinfo{title}{Near-unity photoluminescence quantum yield in
  {MoS}$_2$}.
\newblock \emph{\bibinfo{journal}{Science}} \textbf{\bibinfo{volume}{350}},
  \bibinfo{pages}{1065--1068} (\bibinfo{year}{2015}).
\newblock \urlprefix\url{https://doi.org/10.1126/science.aad2114}.

\bibitem{Cai2015}
\bibinfo{author}{Cai, L.} \emph{et~al.}
\newblock \bibinfo{title}{Vacancy-induced ferromagnetism of {MoS}$_2$
  nanosheets}.
\newblock \emph{\bibinfo{journal}{Journal of the American Chemical Society}}
  \textbf{\bibinfo{volume}{137}}, \bibinfo{pages}{2622--2627}
  (\bibinfo{year}{2015}).
\newblock \urlprefix\url{https://doi.org/10.1021/ja5120908}.

\bibitem{Guguchia2018}
\bibinfo{author}{Guguchia, Z.} \emph{et~al.}
\newblock \bibinfo{title}{Magnetism in semiconducting molybdenum
  dichalcogenides}.
\newblock \emph{\bibinfo{journal}{Science Advances}}
  \textbf{\bibinfo{volume}{4}} (\bibinfo{year}{2018}).
\newblock \urlprefix\url{https://doi.org/10.1126/sciadv.aat3672}.

\bibitem{Mathew.2012}
\bibinfo{author}{Mathew, S.} \emph{et~al.}
\newblock \bibinfo{title}{Magnetism in {MoS}$_2$ induced by proton
  irradiation}.
\newblock \emph{\bibinfo{journal}{Applied Physics Letters}}
  \textbf{\bibinfo{volume}{101}}, \bibinfo{pages}{102103}
  (\bibinfo{year}{2012}).
\newblock \urlprefix\url{https://doi.org/10.1063/1.4750237}.

\bibitem{Yun.2020}
\bibinfo{author}{Yun, S.~J.} \emph{et~al.}
\newblock \bibinfo{title}{Ferromagnetic order at room temperature in monolayer
  {WSe}$_2$ semiconductor via vanadium dopant}.
\newblock \emph{\bibinfo{journal}{Advanced Science}}
  \textbf{\bibinfo{volume}{7}}, \bibinfo{pages}{1903076}
  (\bibinfo{year}{2020}).
\newblock \urlprefix\url{https://doi.org/10.1002/advs.201903076}.

\bibitem{Nguyen.2021}
\bibinfo{author}{Nguyen, L.-A.~T.} \emph{et~al.}
\newblock \bibinfo{title}{Spin-selective hole{\textendash}exciton coupling in a
  {V}-doped {WSe}$_2$ ferromagnetic semiconductor at room temperature}.
\newblock \emph{\bibinfo{journal}{{ACS} Nano}} \textbf{\bibinfo{volume}{15}},
  \bibinfo{pages}{20267--20277} (\bibinfo{year}{2021}).
\newblock \urlprefix\url{https://doi.org/10.1021/acsnano.1c08375}.

\bibitem{Nisi.2022}
\bibinfo{author}{Nisi, K.} \emph{et~al.}
\newblock \bibinfo{title}{Defect-engineered magnetic field dependent
  optoelectronics of vanadium doped tungsten diselenide monolayers}.
\newblock \emph{\bibinfo{journal}{Advanced Optical Materials}}
  \bibinfo{pages}{2102711} (\bibinfo{year}{2022}).
\newblock \urlprefix\url{https://doi.org/10.1002/adom.202102711}.

\bibitem{Lu.2020}
\bibinfo{author}{Lu, X.}, \bibinfo{author}{Fei, R.}, \bibinfo{author}{Zhu, L.}
  \& \bibinfo{author}{Yang, L.}
\newblock \bibinfo{title}{Meron-like topological spin defects in monolayer
  {CrCl}$_3$}.
\newblock \emph{\bibinfo{journal}{Nature Communications}}
  \textbf{\bibinfo{volume}{11}} (\bibinfo{year}{2020}).
\newblock \urlprefix\url{https://doi.org/10.1038/s41467-020-18573-8}.

\bibitem{Graham.2020}
\bibinfo{author}{Graham, J.~N.} \emph{et~al.}
\newblock \bibinfo{title}{Local nuclear and magnetic order in the
  two-dimensional spin glass {Mn}$_{0.5}${Fe}$_{0.5}${PS}$_3$}.
\newblock \emph{\bibinfo{journal}{Physical Review Materials}}
  \textbf{\bibinfo{volume}{4}} (\bibinfo{year}{2020}).
\newblock \urlprefix\url{https://doi.org/10.1103/physrevmaterials.4.084401}.

\bibitem{Beck.2021}
\bibinfo{author}{Beck, R.~A.}, \bibinfo{author}{Lu, L.},
  \bibinfo{author}{Sushko, P.~V.}, \bibinfo{author}{Xu, X.} \&
  \bibinfo{author}{Li, X.}
\newblock \bibinfo{title}{Defect-induced magnetic skyrmion in a two-dimensional
  chromium triiodide monolayer}.
\newblock \emph{\bibinfo{journal}{{JACS} Au}} \textbf{\bibinfo{volume}{1}},
  \bibinfo{pages}{1362--1367} (\bibinfo{year}{2021}).
\newblock \urlprefix\url{https://doi.org/10.1021/jacsau.1c00142}.

\bibitem{Klein.2021}
\bibinfo{author}{Klein, J.} \emph{et~al.}
\newblock \bibinfo{title}{Control of structure and spin texture in the van der
  {Waals} layered magnet {CrSBr}}.
\newblock \emph{\bibinfo{journal}{Nature Communications}}
  \textbf{\bibinfo{volume}{13}} (\bibinfo{year}{2022}).
\newblock \urlprefix\url{https://doi.org/10.1038/s41467-022-32737-8}.

\bibitem{Huang.2017}
\bibinfo{author}{Huang, B.} \emph{et~al.}
\newblock \bibinfo{title}{Layer-dependent ferromagnetism in a van der waals
  crystal down to the monolayer limit}.
\newblock \emph{\bibinfo{journal}{Nature}} \textbf{\bibinfo{volume}{546}},
  \bibinfo{pages}{270--273} (\bibinfo{year}{2017}).
\newblock \urlprefix\url{https://doi.org/10.1038/nature22391}.

\bibitem{Gong.2017}
\bibinfo{author}{Gong, C.} \emph{et~al.}
\newblock \bibinfo{title}{Discovery of intrinsic ferromagnetism in
  two-dimensional van der {Waals} crystals}.
\newblock \emph{\bibinfo{journal}{Nature}} \textbf{\bibinfo{volume}{546}},
  \bibinfo{pages}{265--269} (\bibinfo{year}{2017}).
\newblock \urlprefix\url{https://doi.org/10.1038/nature22060}.

\bibitem{Shcherbakov.2018}
\bibinfo{author}{Shcherbakov, D.} \emph{et~al.}
\newblock \bibinfo{title}{{Raman} spectroscopy, photocatalytic degradation, and
  stabilization of atomically thin chromium tri-iodide}.
\newblock \emph{\bibinfo{journal}{Nano Letters}} \textbf{\bibinfo{volume}{18}},
  \bibinfo{pages}{4214--4219} (\bibinfo{year}{2018}).
\newblock \urlprefix\url{https://doi.org/10.1021/acs.nanolett.8b01131}.

\bibitem{Katscher.1966}
\bibinfo{author}{Katscher, H.} \& \bibinfo{author}{Hahn, H.}
\newblock \bibinfo{title}{{Über Chalkogenidhalogenide des dreiwertigen
  Chroms}}.
\newblock \emph{\bibinfo{journal}{Die Naturwissenschaften}}
  \textbf{\bibinfo{volume}{53}}, \bibinfo{pages}{361--361}
  (\bibinfo{year}{1966}).
\newblock \urlprefix\url{https://doi.org/10.1007/bf00621875}.

\bibitem{Gser.1990}
\bibinfo{author}{G\"{o}ser, O.}, \bibinfo{author}{Paul, W.} \&
  \bibinfo{author}{Kahle, H.}
\newblock \bibinfo{title}{Magnetic properties of {CrSBr}}.
\newblock \emph{\bibinfo{journal}{Journal of Magnetism and Magnetic Materials}}
  \textbf{\bibinfo{volume}{92}}, \bibinfo{pages}{129--136}
  (\bibinfo{year}{1990}).
\newblock \urlprefix\url{https://doi.org/10.1016/0304-8853(90)90689-n}.

\bibitem{Wang.2019}
\bibinfo{author}{Wang, C.} \emph{et~al.}
\newblock \bibinfo{title}{A family of high-temperature ferromagnetic monolayers
  with locked spin-dichroism-mobility anisotropy: {MnNX} and {CrCX} {(X=Cl, Br,
  I; C=S, Se, Te)}}.
\newblock \emph{\bibinfo{journal}{Science Bulletin}}
  \textbf{\bibinfo{volume}{64}}, \bibinfo{pages}{293--300}
  (\bibinfo{year}{2019}).
\newblock \urlprefix\url{https://doi.org/10.1016/j.scib.2019.02.011}.

\bibitem{Telford.2020}
\bibinfo{author}{Telford, E.~J.} \emph{et~al.}
\newblock \bibinfo{title}{Layered antiferromagnetism induces large negative
  magnetoresistance in the van der {Waals} semiconductor {CrSBr}}.
\newblock \emph{\bibinfo{journal}{Advanced Materials}}
  \textbf{\bibinfo{volume}{32}}, \bibinfo{pages}{2003240}
  (\bibinfo{year}{2020}).
\newblock \urlprefix\url{https://doi.org/10.1002/adma.202003240}.

\bibitem{Klein.2022}
\bibinfo{author}{Klein, J.} \emph{et~al.}
\newblock \bibinfo{title}{The bulk van der {Waals} layered magnet {CrSBr} is a
  quasi-{1D} quantum material}.
\newblock \emph{\bibinfo{journal}{arXiv}}  (\bibinfo{year}{2022}).
\newblock \urlprefix\url{https://arxiv.org/abs/2205.13456}.
\newblock \eprint{arXiv:2205.13456}.

\bibitem{Telford.2022}
\bibinfo{author}{Telford, E.~J.} \emph{et~al.}
\newblock \bibinfo{title}{Coupling between magnetic order and charge transport
  in a two-dimensional magnetic semiconductor}.
\newblock \emph{\bibinfo{journal}{Nature Materials}}
  \textbf{\bibinfo{volume}{21}}, \bibinfo{pages}{754--760}
  (\bibinfo{year}{2022}).
\newblock \urlprefix\url{https://doi.org/10.1038/s41563-022-01245-x}.

\bibitem{Lee.2021}
\bibinfo{author}{Lee, K.} \emph{et~al.}
\newblock \bibinfo{title}{Magnetic order and symmetry in the 2d semiconductor
  {CrSBr}}.
\newblock \emph{\bibinfo{journal}{Nano Letters}} \textbf{\bibinfo{volume}{21}},
  \bibinfo{pages}{3511--3517} (\bibinfo{year}{2021}).
\newblock \urlprefix\url{https://doi.org/10.1021/acs.nanolett.1c00219}.

\bibitem{Wilson.2021}
\bibinfo{author}{Wilson, N.~P.} \emph{et~al.}
\newblock \bibinfo{title}{Interlayer electronic coupling on demand in a {2D}
  magnetic semiconductor}.
\newblock \emph{\bibinfo{journal}{Nature Materials}}
  \textbf{\bibinfo{volume}{20}}, \bibinfo{pages}{1657--1662}
  (\bibinfo{year}{2021}).
\newblock \urlprefix\url{https://doi.org/10.1038/s41563-021-01070-8}.

\bibitem{Klein.2022a}
\bibinfo{author}{Klein, J.} \emph{et~al.}
\newblock \bibinfo{title}{Sensing the local magnetic environment through
  optically active defects in a layered magnetic semiconductor}.
\newblock \emph{\bibinfo{journal}{arXiv}}  (\bibinfo{year}{2022}).
\newblock \urlprefix\url{https://arxiv.org/abs/2207.02884}.
\newblock \eprint{arXiv:2207.02884}.

\bibitem{Wu.2022}
\bibinfo{author}{Wu, F.} \emph{et~al.}
\newblock \bibinfo{title}{{Quasi 1D Electronic Transport in a 2D Magnetic
  Semiconductor}}.
\newblock \emph{\bibinfo{journal}{Advanced Materials}} \bibinfo{pages}{2109759}
  (\bibinfo{year}{2022}).
\newblock \urlprefix\url{https://doi.org/10.1002/adma.202109759}.

\bibitem{Boix-Constant.2022}
\bibinfo{author}{Boix-Constant, C.} \emph{et~al.}
\newblock \bibinfo{title}{Probing the spin dimensionality in single-layer
  {CrSBr} van der {Waals} heterostructures by magneto-transport measurements}.
\newblock \emph{\bibinfo{journal}{Advanced Materials}}
  \textbf{\bibinfo{volume}{34}}, \bibinfo{pages}{2204940}
  (\bibinfo{year}{2022}).
\newblock \urlprefix\url{https://doi.org/10.1002/adma.202204940}.

\bibitem{Paz.2022}
\bibinfo{author}{L{\'{o}}pez-Paz, S.~A.} \emph{et~al.}
\newblock \bibinfo{title}{Dynamic magnetic crossover at the origin of the
  hidden-order in van der {Waals} antiferromagnet {CrSBr}}.
\newblock \emph{\bibinfo{journal}{Nature Communications}}
  \textbf{\bibinfo{volume}{13}} (\bibinfo{year}{2022}).
\newblock \urlprefix\url{https://doi.org/10.1038/s41467-022-32290-4}.

\bibitem{Ye2022}
\bibinfo{author}{Ye, C.} \emph{et~al.}
\newblock \bibinfo{title}{Layer-dependent interlayer antiferromagnetic spin
  reorientation in air-stable semiconductor {CrSBr}}.
\newblock \emph{\bibinfo{journal}{{ACS} Nano}}  (\bibinfo{year}{2022}).
\newblock \urlprefix\url{https://doi.org/10.1021/acsnano.2c01151}.

\bibitem{Malard.2009}
\bibinfo{author}{Malard, L.}, \bibinfo{author}{Pimenta, M.},
  \bibinfo{author}{Dresselhaus, G.} \& \bibinfo{author}{Dresselhaus, M.}
\newblock \bibinfo{title}{{Raman} spectroscopy in graphene}.
\newblock \emph{\bibinfo{journal}{Physics Reports}}
  \textbf{\bibinfo{volume}{473}}, \bibinfo{pages}{51--87}
  (\bibinfo{year}{2009}).
\newblock \urlprefix\url{https://doi.org/10.1016/j.physrep.2009.02.003}.

\bibitem{Herziger.2014}
\bibinfo{author}{Herziger, F.}, \bibinfo{author}{Tyborski, C.},
  \bibinfo{author}{Ochedowski, O.}, \bibinfo{author}{Schleberger, M.} \&
  \bibinfo{author}{Maultzsch, J.}
\newblock \bibinfo{title}{Double-resonant {LA} phonon scattering in defective
  graphene and carbon nanotubes}.
\newblock \emph{\bibinfo{journal}{Physical Review B}}
  \textbf{\bibinfo{volume}{90}} (\bibinfo{year}{2014}).
\newblock \urlprefix\url{https://doi.org/10.1103/physrevb.90.245431}.

\bibitem{Cenker2022}
\bibinfo{author}{Cenker, J.} \emph{et~al.}
\newblock \bibinfo{title}{Reversible strain-induced magnetic phase transition
  in a van der {Waals} magnet}.
\newblock \emph{\bibinfo{journal}{Nature Nanotechnology}}
  (\bibinfo{year}{2022}).
\newblock \urlprefix\url{https://doi.org/10.1038/s41565-021-01052-6}.

\bibitem{Bae.2022}
\bibinfo{author}{Bae, Y.~J.} \emph{et~al.}
\newblock \bibinfo{title}{Exciton-coupled coherent magnons in a $2d$
  semiconductor}.
\newblock \emph{\bibinfo{journal}{Nature}} \textbf{\bibinfo{volume}{609}},
  \bibinfo{pages}{282--286} (\bibinfo{year}{2022}).
\newblock \urlprefix\url{https://doi.org/10.1038/s41586-022-05024-1}.

\bibitem{Loudon.1964}
\bibinfo{author}{Loudon, R.}
\newblock \bibinfo{title}{The {Raman} effect in crystals}.
\newblock \emph{\bibinfo{journal}{Advances in Physics}}
  \textbf{\bibinfo{volume}{13}}, \bibinfo{pages}{423--482}
  (\bibinfo{year}{1964}).
\newblock \urlprefix\url{https://doi.org/10.1080/00018736400101051}.

\bibitem{Fausti.2007}
\bibinfo{author}{Fausti, D.} \emph{et~al.}
\newblock \bibinfo{title}{Symmetry disquisition on {the TiOX phase} diagram(x =
  br, cl)}.
\newblock \emph{\bibinfo{journal}{Physical Review B}}
  \textbf{\bibinfo{volume}{75}} (\bibinfo{year}{2007}).
\newblock \urlprefix\url{https://doi.org/10.1103/physrevb.75.245114}.

\bibitem{Bykov.2013}
\bibinfo{author}{Bykov, M.} \emph{et~al.}
\newblock \bibinfo{title}{High-pressure behavior of {FeOCl}}.
\newblock \emph{\bibinfo{journal}{Physical Review B}}
  \textbf{\bibinfo{volume}{88}} (\bibinfo{year}{2013}).
\newblock \urlprefix\url{https://doi.org/10.1103/physrevb.88.014110}.

\bibitem{Zhang2019}
\bibinfo{author}{Zhang, T.} \emph{et~al.}
\newblock \bibinfo{title}{Magnetism and optical anisotropy in van der {Waals}
  antiferromagnetic insulator {CrOCl}}.
\newblock \emph{\bibinfo{journal}{{ACS} Nano}} \textbf{\bibinfo{volume}{13}},
  \bibinfo{pages}{11353--11362} (\bibinfo{year}{2019}).
\newblock \urlprefix\url{https://doi.org/10.1021/acsnano.9b04726}.

\bibitem{Lee2010}
\bibinfo{author}{Lee, C.} \emph{et~al.}
\newblock \bibinfo{title}{Anomalous lattice vibrations of single- and few-layer
  {MoS}$_2$}.
\newblock \emph{\bibinfo{journal}{{ACS} Nano}} \textbf{\bibinfo{volume}{4}},
  \bibinfo{pages}{2695--2700} (\bibinfo{year}{2010}).
\newblock \urlprefix\url{https://doi.org/10.1021/nn1003937}.

\bibitem{MolinaSnchez2011}
\bibinfo{author}{Molina-S{\'{a}}nchez, A.} \& \bibinfo{author}{Wirtz, L.}
\newblock \bibinfo{title}{Phonons in single-layer and few-layer {MoS}$_2$}.
\newblock \emph{\bibinfo{journal}{Physical Review B}}
  \textbf{\bibinfo{volume}{84}} (\bibinfo{year}{2011}).
\newblock \urlprefix\url{https://doi.org/10.1103/physrevb.84.155413}.

\bibitem{Bell2009}
\bibinfo{author}{Bell, D.~C.}, \bibinfo{author}{Lemme, M.~C.},
  \bibinfo{author}{Stern, L.~A.}, \bibinfo{author}{Williams, J.~R.} \&
  \bibinfo{author}{Marcus, C.~M.}
\newblock \bibinfo{title}{Precision cutting and patterning of graphene with
  helium ions}.
\newblock \emph{\bibinfo{journal}{Nanotechnology}}
  \textbf{\bibinfo{volume}{20}}, \bibinfo{pages}{455301}
  (\bibinfo{year}{2009}).
\newblock \urlprefix\url{https://doi.org/10.1088/0957-4484/20/45/455301}.

\bibitem{Fox2013}
\bibinfo{author}{Fox, D.} \emph{et~al.}
\newblock \bibinfo{title}{Helium ion microscopy of graphene: beam damage, image
  quality and edge contrast}.
\newblock \emph{\bibinfo{journal}{Nanotechnology}}
  \textbf{\bibinfo{volume}{24}}, \bibinfo{pages}{335702}
  (\bibinfo{year}{2013}).
\newblock \urlprefix\url{https://doi.org/10.1088/0957-4484/24/33/335702}.

\bibitem{Fox2015}
\bibinfo{author}{Fox, D.~S.} \emph{et~al.}
\newblock \bibinfo{title}{Nanopatterning and electrical tuning of {MoS}$_2$
  layers with a subnanometer helium ion beam}.
\newblock \emph{\bibinfo{journal}{Nano Letters}} \textbf{\bibinfo{volume}{15}},
  \bibinfo{pages}{5307--5313} (\bibinfo{year}{2015}).
\newblock \urlprefix\url{https://doi.org/10.1021/acs.nanolett.5b01673}.

\bibitem{Jadwiszczak2019}
\bibinfo{author}{Jadwiszczak, J.} \emph{et~al.}
\newblock \bibinfo{title}{{MoS}$_2$ memtransistors fabricated by localized
  helium ion beam irradiation}.
\newblock \emph{\bibinfo{journal}{{ACS} Nano}} \textbf{\bibinfo{volume}{13}},
  \bibinfo{pages}{14262--14273} (\bibinfo{year}{2019}).
\newblock \urlprefix\url{https://doi.org/10.1021/acsnano.9b07421}.

\bibitem{Guo2015}
\bibinfo{author}{Guo, Y.} \emph{et~al.}
\newblock \bibinfo{title}{Charge trapping at the {MoS}$_2$-{SiO}$_2$ interface
  and its effects on the characteristics of {MoS}$_2$ metal-oxide-semiconductor
  field effect transistors}.
\newblock \emph{\bibinfo{journal}{Applied Physics Letters}}
  \textbf{\bibinfo{volume}{106}}, \bibinfo{pages}{103109}
  (\bibinfo{year}{2015}).
\newblock \urlprefix\url{https://doi.org/10.1063/1.4914968}.

\bibitem{Chae2017}
\bibinfo{author}{Chae, W.~H.}, \bibinfo{author}{Cain, J.~D.},
  \bibinfo{author}{Hanson, E.~D.}, \bibinfo{author}{Murthy, A.~A.} \&
  \bibinfo{author}{Dravid, V.~P.}
\newblock \bibinfo{title}{Substrate-induced strain and charge doping in
  {CVD}-grown monolayer {MoS}$_2$}.
\newblock \emph{\bibinfo{journal}{Applied Physics Letters}}
  \textbf{\bibinfo{volume}{111}}, \bibinfo{pages}{143106}
  (\bibinfo{year}{2017}).
\newblock \urlprefix\url{https://doi.org/10.1063/1.4998284}.

\bibitem{Romero.2008}
\bibinfo{author}{Romero, H.~E.} \emph{et~al.}
\newblock \bibinfo{title}{n-type behavior of graphene supported on si/{SiO}$_2$
  substrates}.
\newblock \emph{\bibinfo{journal}{{ACS} Nano}} \textbf{\bibinfo{volume}{2}},
  \bibinfo{pages}{2037--2044} (\bibinfo{year}{2008}).
\newblock \urlprefix\url{https://doi.org/10.1021/nn800354m}.

\bibitem{Ziegler.2010}
\bibinfo{author}{Ziegler, J.~F.}, \bibinfo{author}{Ziegler, M.} \&
  \bibinfo{author}{Biersack, J.}
\newblock \bibinfo{title}{{SRIM} {\textendash} the stopping and range of ions
  in matter (2010)}.
\newblock \emph{\bibinfo{journal}{Nuclear Instruments and Methods in Physics
  Research Section B: Beam Interactions with Materials and Atoms}}
  \textbf{\bibinfo{volume}{268}}, \bibinfo{pages}{1818--1823}
  (\bibinfo{year}{2010}).
\newblock \urlprefix\url{https://doi.org/10.1016/j.nimb.2010.02.091}.

\bibitem{Zhang.2022}
\bibinfo{author}{Zhang, T.} \emph{et~al.}
\newblock \bibinfo{title}{Degradation chemistry and kinetic stabilization of
  magnetic {CrI}$_3$}.
\newblock \emph{\bibinfo{journal}{Journal of the American Chemical Society}}
  \textbf{\bibinfo{volume}{144}}, \bibinfo{pages}{5295--5303}
  (\bibinfo{year}{2022}).
\newblock \urlprefix\url{https://doi.org/10.1021/jacs.1c08906}.

\bibitem{Campbell1986}
\bibinfo{author}{Campbell, I.} \& \bibinfo{author}{Fauchet, P.}
\newblock \bibinfo{title}{The effects of microcrystal size and shape on the one
  phonon {Raman} spectra of crystalline semiconductors}.
\newblock \emph{\bibinfo{journal}{Solid State Communications}}
  \textbf{\bibinfo{volume}{58}}, \bibinfo{pages}{739--741}
  (\bibinfo{year}{1986}).
\newblock \urlprefix\url{https://doi.org/10.1016/0038-1098(86)90513-2}.

\bibitem{Richter1981}
\bibinfo{author}{Richter, H.}, \bibinfo{author}{Wang, Z.} \&
  \bibinfo{author}{Ley, L.}
\newblock \bibinfo{title}{The one phonon {Raman} spectrum in microcrystalline
  silicon}.
\newblock \emph{\bibinfo{journal}{Solid State Communications}}
  \textbf{\bibinfo{volume}{39}}, \bibinfo{pages}{625--629}
  (\bibinfo{year}{1981}).
\newblock \urlprefix\url{https://doi.org/10.1016/0038-1098(81)90337-9}.

\bibitem{Mignuzzi2015}
\bibinfo{author}{Mignuzzi, S.} \emph{et~al.}
\newblock \bibinfo{title}{Effect of disorder on {Raman} scattering of
  single-{layer MoS}$_2$}.
\newblock \emph{\bibinfo{journal}{Physical Review B}}
  \textbf{\bibinfo{volume}{91}} (\bibinfo{year}{2015}).
\newblock \urlprefix\url{https://doi.org/10.1103/physrevb.91.195411}.

\bibitem{Shi2016}
\bibinfo{author}{Shi, W.} \emph{et~al.}
\newblock \bibinfo{title}{Phonon confinement effect in two-dimensional
  nanocrystallites of monolayer {MoS}$_2$ to probe phonon dispersion trends
  away from brillouin-zone center}.
\newblock \emph{\bibinfo{journal}{Chinese Physics Letters}}
  \textbf{\bibinfo{volume}{33}}, \bibinfo{pages}{057801}
  (\bibinfo{year}{2016}).
\newblock \urlprefix\url{https://doi.org/10.1088/0256-307x/33/5/057801}.

\bibitem{MartinsFerreira2010}
\bibinfo{author}{Ferreira, E. H.~M.} \emph{et~al.}
\newblock \bibinfo{title}{Evolution of the {Raman} spectra from single-, few-,
  and many-layer graphene with increasing disorder}.
\newblock \emph{\bibinfo{journal}{Physical Review B}}
  \textbf{\bibinfo{volume}{82}} (\bibinfo{year}{2010}).
\newblock \urlprefix\url{https://doi.org/10.1103/physrevb.82.125429}.

\bibitem{Wang.2020}
\bibinfo{author}{Wang, H.}, \bibinfo{author}{Qi, J.} \& \bibinfo{author}{Qian,
  X.}
\newblock \bibinfo{title}{Electrically tunable high curie temperature
  two-dimensional ferromagnetism in van der {Waals} layered crystals}.
\newblock \emph{\bibinfo{journal}{Applied Physics Letters}}
  \textbf{\bibinfo{volume}{117}}, \bibinfo{pages}{083102}
  (\bibinfo{year}{2020}).
\newblock \urlprefix\url{https://doi.org/10.1063/5.0014865}.

\bibitem{Arora.1987}
\bibinfo{author}{Arora, A.~K.} \& \bibinfo{author}{Ramdas, A.~K.}
\newblock \bibinfo{title}{Resonance {Raman} scattering from defects in {CdSe}}.
\newblock \emph{\bibinfo{journal}{Physical Review B}}
  \textbf{\bibinfo{volume}{35}}, \bibinfo{pages}{4345--4350}
  (\bibinfo{year}{1987}).
\newblock \urlprefix\url{https://doi.org/10.1103/physrevb.35.4345}.

\bibitem{Munz.1980}
\bibinfo{author}{Munz, D.} \& \bibinfo{author}{Pilkuhn, M.}
\newblock \bibinfo{title}{Resonant {Raman} studies of bound excitons in {CdS}}.
\newblock \emph{\bibinfo{journal}{Solid State Communications}}
  \textbf{\bibinfo{volume}{36}}, \bibinfo{pages}{205--209}
  (\bibinfo{year}{1980}).
\newblock \urlprefix\url{https://doi.org/10.1016/0038-1098(80)90261-6}.

\bibitem{Berg.1986}
\bibinfo{author}{Berg, R.~S.} \& \bibinfo{author}{Yu, P.~Y.}
\newblock \bibinfo{title}{Enhancement of defect-induced {Raman} modes at the
  fundamental absorption edge of electron-irradiated {GaAs}}.
\newblock \emph{\bibinfo{journal}{Physical Review B}}
  \textbf{\bibinfo{volume}{33}}, \bibinfo{pages}{7349--7352}
  (\bibinfo{year}{1986}).
\newblock \urlprefix\url{https://doi.org/10.1103/physrevb.33.7349}.

\bibitem{Granado1999}
\bibinfo{author}{Granado, E.} \emph{et~al.}
\newblock \bibinfo{title}{Magnetic ordering effects in the {Raman} spectra of
  magnetic ordering effects in the {Raman} spectra of
  {La$_{1-x}$Mn$_{1-x}$O$_3$}}.
\newblock \emph{\bibinfo{journal}{Physical Review B}}
  \textbf{\bibinfo{volume}{60}}, \bibinfo{pages}{11879--11882}
  (\bibinfo{year}{1999}).
\newblock \urlprefix\url{https://doi.org/10.1103/physrevb.60.11879}.

\bibitem{Laverdire.2006}
\bibinfo{author}{Laverdi{\`{e}}re, J.} \emph{et~al.}
\newblock \bibinfo{title}{Spin-phonon coupling in orthorhombic {RMnO}$_3$ ({R =
  Pr, Nd, Sm, Eu, Gd, Tb, Dy, Ho, Y}): A {Raman} study}.
\newblock \emph{\bibinfo{journal}{Physical Review B}}
  \textbf{\bibinfo{volume}{73}} (\bibinfo{year}{2006}).
\newblock \urlprefix\url{https://doi.org/10.1103/physrevb.73.214301}.

\bibitem{Xu.2022}
\bibinfo{author}{Xu, X.} \emph{et~al.}
\newblock \bibinfo{title}{Strong spin-phonon coupling in two-dimensional
  magnetic semiconductor {CrSBr}}.
\newblock \emph{\bibinfo{journal}{The Journal of Physical Chemistry C}}
  \textbf{\bibinfo{volume}{126}}, \bibinfo{pages}{10574--10583}
  (\bibinfo{year}{2022}).
\newblock \urlprefix\url{https://doi.org/10.1021/acs.jpcc.2c02742}.

\bibitem{Du.2019}
\bibinfo{author}{Du, L.} \emph{et~al.}
\newblock \bibinfo{title}{Lattice dynamics, phonon chirality, and
  spin{\textendash}phonon coupling in {2D} itinerant ferromagnet
  {Fe}$_3${GeTe}$_2$}.
\newblock \emph{\bibinfo{journal}{Advanced Functional Materials}}
  \textbf{\bibinfo{volume}{29}}, \bibinfo{pages}{1904734}
  (\bibinfo{year}{2019}).
\newblock \urlprefix\url{https://doi.org/10.1002/adfm.201904734}.

\bibitem{Suzuki.1973}
\bibinfo{author}{Suzuki, N.} \& \bibinfo{author}{Kamimura, H.}
\newblock \bibinfo{title}{Theory of spin-dependent phonon {Raman} scattering in
  magnetic crystals}.
\newblock \emph{\bibinfo{journal}{Journal of the Physical Society of Japan}}
  \textbf{\bibinfo{volume}{35}}, \bibinfo{pages}{985--995}
  (\bibinfo{year}{1973}).
\newblock \urlprefix\url{https://doi.org/10.1143/jpsj.35.985}.

\bibitem{Schneider.2010}
\bibinfo{author}{Schneider, G.~F.}, \bibinfo{author}{Calado, V.~E.},
  \bibinfo{author}{Zandbergen, H.}, \bibinfo{author}{Vandersypen, L. M.~K.} \&
  \bibinfo{author}{Dekker, C.}
\newblock \bibinfo{title}{Wedging transfer of nanostructures}.
\newblock \emph{\bibinfo{journal}{Nano Letters}} \textbf{\bibinfo{volume}{10}},
  \bibinfo{pages}{1912--1916} (\bibinfo{year}{2010}).
\newblock \urlprefix\url{https://doi.org/10.1021/nl1008037}.

\bibitem{crystal17}
\bibinfo{author}{Dovesi, R.} \emph{et~al.}
\newblock \bibinfo{title}{Quantum-mechanical condensed matter simulations with
  crystal}.
\newblock \emph{\bibinfo{journal}{WIRE: Comp. Mol. Sci.}}
  \textbf{\bibinfo{volume}{8}}, \bibinfo{pages}{e1360} (\bibinfo{year}{2018}).
\newblock \urlprefix\url{https://doi.org/10.1002/wcms.1360}.

\bibitem{PBE0}
\bibinfo{author}{Adamo, C.} \& \bibinfo{author}{Barone, V.}
\newblock \bibinfo{title}{Toward reliable density functional methods without
  adjustable parameters: The {PBE}0 model}.
\newblock \emph{\bibinfo{journal}{The Journal of Chemical Physics}}
  \textbf{\bibinfo{volume}{110}}, \bibinfo{pages}{6158--6170}
  (\bibinfo{year}{1999}).
\newblock \urlprefix\url{https://doi.org/10.1063/1.478522}.

\bibitem{pob_basis}
\bibinfo{author}{Peintinger, M.~F.}, \bibinfo{author}{Oliveira, D.~V.} \&
  \bibinfo{author}{Bredow, T.}
\newblock \bibinfo{title}{Consistent gaussian basis sets of triple-zeta valence
  with polarization quality for solid-state calculations}.
\newblock \emph{\bibinfo{journal}{J. Comp. Chem.}}
  \textbf{\bibinfo{volume}{34}}, \bibinfo{pages}{451--459}
  (\bibinfo{year}{2013}).
\newblock \urlprefix\url{https://doi.org/10.1002/jcc.23153}.

\bibitem{maschio2013ab}
\bibinfo{author}{Maschio, L.}, \bibinfo{author}{Kirtman, B.},
  \bibinfo{author}{R{\'e}rat, M.}, \bibinfo{author}{Orlando, R.} \&
  \bibinfo{author}{Dovesi, R.}
\newblock \bibinfo{title}{Ab initio analytical {Raman} intensities for periodic
  systems through a coupled perturbed {Hartree-Fock/Kohn-Sham} method in an
  atomic orbital basis. i. theory}.
\newblock \emph{\bibinfo{journal}{J. Chem. Phys.}}
  \textbf{\bibinfo{volume}{139}}, \bibinfo{pages}{164101}
  (\bibinfo{year}{2013}).
\newblock \urlprefix\url{https://doi.org/10.1063/1.4824442}.

\end{thebibliography}


\begin{thebibliography}{6}
\expandafter\ifx\csname natexlab\endcsname\relax\def\natexlab#1{#1}\fi
\expandafter\ifx\csname bibnamefont\endcsname\relax
  \def\bibnamefont#1{#1}\fi
\expandafter\ifx\csname bibfnamefont\endcsname\relax
  \def\bibfnamefont#1{#1}\fi
\expandafter\ifx\csname citenamefont\endcsname\relax
  \def\citenamefont#1{#1}\fi
\expandafter\ifx\csname url\endcsname\relax
  \def\url#1{\texttt{#1}}\fi
\expandafter\ifx\csname urlprefix\endcsname\relax\def\urlprefix{URL }\fi
\providecommand{\bibinfo}[2]{#2}
\providecommand{\eprint}[2][]{\url{#2}}

\bibitem[{\citenamefont{Lee et~al.}(2021)\citenamefont{Lee, Dismukes, Telford,
  Wiscons, Wang, Xu, Nuckolls, Dean, Roy, and Zhu}}]{Lee.2021}
\bibinfo{author}{\bibfnamefont{K.}~\bibnamefont{Lee}},
  \bibinfo{author}{\bibfnamefont{A.~H.} \bibnamefont{Dismukes}},
  \bibinfo{author}{\bibfnamefont{E.~J.} \bibnamefont{Telford}},
  \bibinfo{author}{\bibfnamefont{R.~A.} \bibnamefont{Wiscons}},
  \bibinfo{author}{\bibfnamefont{J.}~\bibnamefont{Wang}},
  \bibinfo{author}{\bibfnamefont{X.}~\bibnamefont{Xu}},
  \bibinfo{author}{\bibfnamefont{C.}~\bibnamefont{Nuckolls}},
  \bibinfo{author}{\bibfnamefont{C.~R.} \bibnamefont{Dean}},
  \bibinfo{author}{\bibfnamefont{X.}~\bibnamefont{Roy}}, \bibnamefont{and}
  \bibinfo{author}{\bibfnamefont{X.}~\bibnamefont{Zhu}}, \bibinfo{journal}{Nano
  Letters} \textbf{\bibinfo{volume}{21}}, \bibinfo{pages}{3511}
  (\bibinfo{year}{2021}),
  \urlprefix\url{https://doi.org/10.1021/acs.nanolett.1c00219}.

\bibitem[{\citenamefont{Palleschi et~al.}(2020)\citenamefont{Palleschi,
  D'Olimpio, Benassi, Nardone, Alfonsetti, Moccia, Renzelli, Cacioppo, Hichri,
  Jaziri et~al.}}]{Palleschi2020}
\bibinfo{author}{\bibfnamefont{S.}~\bibnamefont{Palleschi}},
  \bibinfo{author}{\bibfnamefont{G.}~\bibnamefont{D'Olimpio}},
  \bibinfo{author}{\bibfnamefont{P.}~\bibnamefont{Benassi}},
  \bibinfo{author}{\bibfnamefont{M.}~\bibnamefont{Nardone}},
  \bibinfo{author}{\bibfnamefont{R.}~\bibnamefont{Alfonsetti}},
  \bibinfo{author}{\bibfnamefont{G.}~\bibnamefont{Moccia}},
  \bibinfo{author}{\bibfnamefont{M.}~\bibnamefont{Renzelli}},
  \bibinfo{author}{\bibfnamefont{O.~A.} \bibnamefont{Cacioppo}},
  \bibinfo{author}{\bibfnamefont{A.}~\bibnamefont{Hichri}},
  \bibinfo{author}{\bibfnamefont{S.}~\bibnamefont{Jaziri}},
  \bibnamefont{et~al.}, \bibinfo{journal}{2D Materials}
  \textbf{\bibinfo{volume}{7}}, \bibinfo{pages}{025001} (\bibinfo{year}{2020}),
  \urlprefix\url{https://doi.org/10.1088/2053-1583/ab5bf8}.

\bibitem[{\citenamefont{Klein et~al.}(2022)\citenamefont{Klein, Song, Pingault,
  Dirnberger, Chi, Curtis, Dana, Bushati, Quan, Dekanovsky
  et~al.}}]{Klein.2022a}
\bibinfo{author}{\bibfnamefont{J.}~\bibnamefont{Klein}},
  \bibinfo{author}{\bibfnamefont{Z.}~\bibnamefont{Song}},
  \bibinfo{author}{\bibfnamefont{B.}~\bibnamefont{Pingault}},
  \bibinfo{author}{\bibfnamefont{F.}~\bibnamefont{Dirnberger}},
  \bibinfo{author}{\bibfnamefont{H.}~\bibnamefont{Chi}},
  \bibinfo{author}{\bibfnamefont{J.~B.} \bibnamefont{Curtis}},
  \bibinfo{author}{\bibfnamefont{R.}~\bibnamefont{Dana}},
  \bibinfo{author}{\bibfnamefont{R.}~\bibnamefont{Bushati}},
  \bibinfo{author}{\bibfnamefont{J.}~\bibnamefont{Quan}},
  \bibinfo{author}{\bibfnamefont{L.}~\bibnamefont{Dekanovsky}},
  \bibnamefont{et~al.}, \bibinfo{journal}{arXiv}  (\bibinfo{year}{2022}),
  \eprint{arXiv:2207.02884}, \urlprefix\url{https://arxiv.org/abs/2207.02884}.

\bibitem[{\citenamefont{Pet{\H{o}} et~al.}(2018)\citenamefont{Pet{\H{o}},
  Oll{\'{a}}r, Vancs{\'{o}}, Popov, Magda, Dobrik, Hwang, Sorokin, and
  Tapaszt{\'{o}}}}]{Pet2018}
\bibinfo{author}{\bibfnamefont{J.}~\bibnamefont{Pet{\H{o}}}},
  \bibinfo{author}{\bibfnamefont{T.}~\bibnamefont{Oll{\'{a}}r}},
  \bibinfo{author}{\bibfnamefont{P.}~\bibnamefont{Vancs{\'{o}}}},
  \bibinfo{author}{\bibfnamefont{Z.~I.} \bibnamefont{Popov}},
  \bibinfo{author}{\bibfnamefont{G.~Z.} \bibnamefont{Magda}},
  \bibinfo{author}{\bibfnamefont{G.}~\bibnamefont{Dobrik}},
  \bibinfo{author}{\bibfnamefont{C.}~\bibnamefont{Hwang}},
  \bibinfo{author}{\bibfnamefont{P.~B.} \bibnamefont{Sorokin}},
  \bibnamefont{and}
  \bibinfo{author}{\bibfnamefont{L.}~\bibnamefont{Tapaszt{\'{o}}}},
  \bibinfo{journal}{Nature Chemistry} \textbf{\bibinfo{volume}{10}},
  \bibinfo{pages}{1246} (\bibinfo{year}{2018}),
  \urlprefix\url{https://doi.org/10.1038/s41557-018-0136-2}.

\bibitem[{\citenamefont{Barja et~al.}(2019)\citenamefont{Barja,
  Refaely-Abramson, Schuler, Qiu, Pulkin, Wickenburg, Ryu, Ugeda, Kastl, Chen
  et~al.}}]{Barja2019}
\bibinfo{author}{\bibfnamefont{S.}~\bibnamefont{Barja}},
  \bibinfo{author}{\bibfnamefont{S.}~\bibnamefont{Refaely-Abramson}},
  \bibinfo{author}{\bibfnamefont{B.}~\bibnamefont{Schuler}},
  \bibinfo{author}{\bibfnamefont{D.~Y.} \bibnamefont{Qiu}},
  \bibinfo{author}{\bibfnamefont{A.}~\bibnamefont{Pulkin}},
  \bibinfo{author}{\bibfnamefont{S.}~\bibnamefont{Wickenburg}},
  \bibinfo{author}{\bibfnamefont{H.}~\bibnamefont{Ryu}},
  \bibinfo{author}{\bibfnamefont{M.~M.} \bibnamefont{Ugeda}},
  \bibinfo{author}{\bibfnamefont{C.}~\bibnamefont{Kastl}},
  \bibinfo{author}{\bibfnamefont{C.}~\bibnamefont{Chen}}, \bibnamefont{et~al.},
  \bibinfo{journal}{Nature Communications} \textbf{\bibinfo{volume}{10}}
  (\bibinfo{year}{2019}),
  \urlprefix\url{https://doi.org/10.1038/s41467-019-11342-2}.

\bibitem[{\citenamefont{Schuler et~al.}(2019)\citenamefont{Schuler, Qiu,
  Refaely-Abramson, Kastl, Chen, Barja, Koch, Ogletree, Aloni, Schwartzberg
  et~al.}}]{Schuler.2019}
\bibinfo{author}{\bibfnamefont{B.}~\bibnamefont{Schuler}},
  \bibinfo{author}{\bibfnamefont{D.~Y.} \bibnamefont{Qiu}},
  \bibinfo{author}{\bibfnamefont{S.}~\bibnamefont{Refaely-Abramson}},
  \bibinfo{author}{\bibfnamefont{C.}~\bibnamefont{Kastl}},
  \bibinfo{author}{\bibfnamefont{C.~T.} \bibnamefont{Chen}},
  \bibinfo{author}{\bibfnamefont{S.}~\bibnamefont{Barja}},
  \bibinfo{author}{\bibfnamefont{R.~J.} \bibnamefont{Koch}},
  \bibinfo{author}{\bibfnamefont{D.~F.} \bibnamefont{Ogletree}},
  \bibinfo{author}{\bibfnamefont{S.}~\bibnamefont{Aloni}},
  \bibinfo{author}{\bibfnamefont{A.~M.} \bibnamefont{Schwartzberg}},
  \bibnamefont{et~al.}, \bibinfo{journal}{Physical Review Letters}
  \textbf{\bibinfo{volume}{123}} (\bibinfo{year}{2019}),
  \urlprefix\url{https://doi.org/10.1103/physrevlett.123.076801}.

\end{thebibliography}

%
%###############################################################################
%								BIBLIOGRAPHY
%##############################################################################
%

\end{document}

%############################## TITLE #########################################
\title{Supporting Information - Probing defects and spin-phonon coupling in CrSBr via resonant Raman scattering}
%##############################################################################
%
%############################ AUTHORS #########################################
\author{K.~Torres}\email{kierstin@mit.edu}
\affiliation{Department of Materials Science and Engineering, Massachusetts Institute of Technology, Cambridge, Massachusetts 02139, USA}
\author{A.~Kuc}%\email{a.kuc@hzdr.de}
\affiliation{Helmholtz-Zentrum Dresden-Rossendorf, Abteilung Ressourcen\"okologie, Forschungsstelle Leipzig, Permoserstr. 15, 04318 Leipzig, Germany}
\author{L.~Maschio}
\affiliation{Dipartimento di Chimica and NIS Centre of Excellence, Universit\`a di Torino, via P. Giuria 5, I-10125 Turin, Italy}
\author{T.~Pham}
\affiliation{Department of Materials Science and Engineering, Massachusetts Institute of Technology, Cambridge, Massachusetts 02139, USA}
\author{K.~Reidy}
\affiliation{Department of Materials Science and Engineering, Massachusetts Institute of Technology, Cambridge, Massachusetts 02139, USA}
%
%\author{J.~Luxa}
%\affiliation{Department of Inorganic Chemistry, University of Chemistry and Technology Prague, Technická 5, 166 28 Prague 6, Czech Republic}
%
\author{L.~Dekanovsky}%\email{dekanovl@vscht.cz}
\affiliation{Department of Inorganic Chemistry, University of Chemistry and Technology Prague, Technická 5, 166 28 Prague 6, Czech Republic}
\author{Z.~Sofer}
\affiliation{Department of Inorganic Chemistry, University of Chemistry and Technology Prague, Technická 5, 166 28 Prague 6, Czech Republic}
\author{F.~M.~Ross}\email{fmross@mit.edu}
\affiliation{Department of Materials Science and Engineering, Massachusetts Institute of Technology, Cambridge, Massachusetts 02139, USA}
\author{J.~Klein}\email{jpklein@mit.edu}
\affiliation{Department of Materials Science and Engineering, Massachusetts Institute of Technology, Cambridge, Massachusetts 02139, USA}
%
%##############################################################################
%
\date{\today}
%
%##############################################################################

%##############################################################################
%
\maketitle
%
%###############################################################################
%								MAIN TEXT
%###############################################################################
%

\tableofcontents

\newpage

%\sectionfont{\titlecap}

\section{Determination of flake thickness.}

%
%###################### Figure ###############################################
\begin{figure*}[ht]
\scalebox{\figurescale}{\includegraphics[width=1\linewidth]{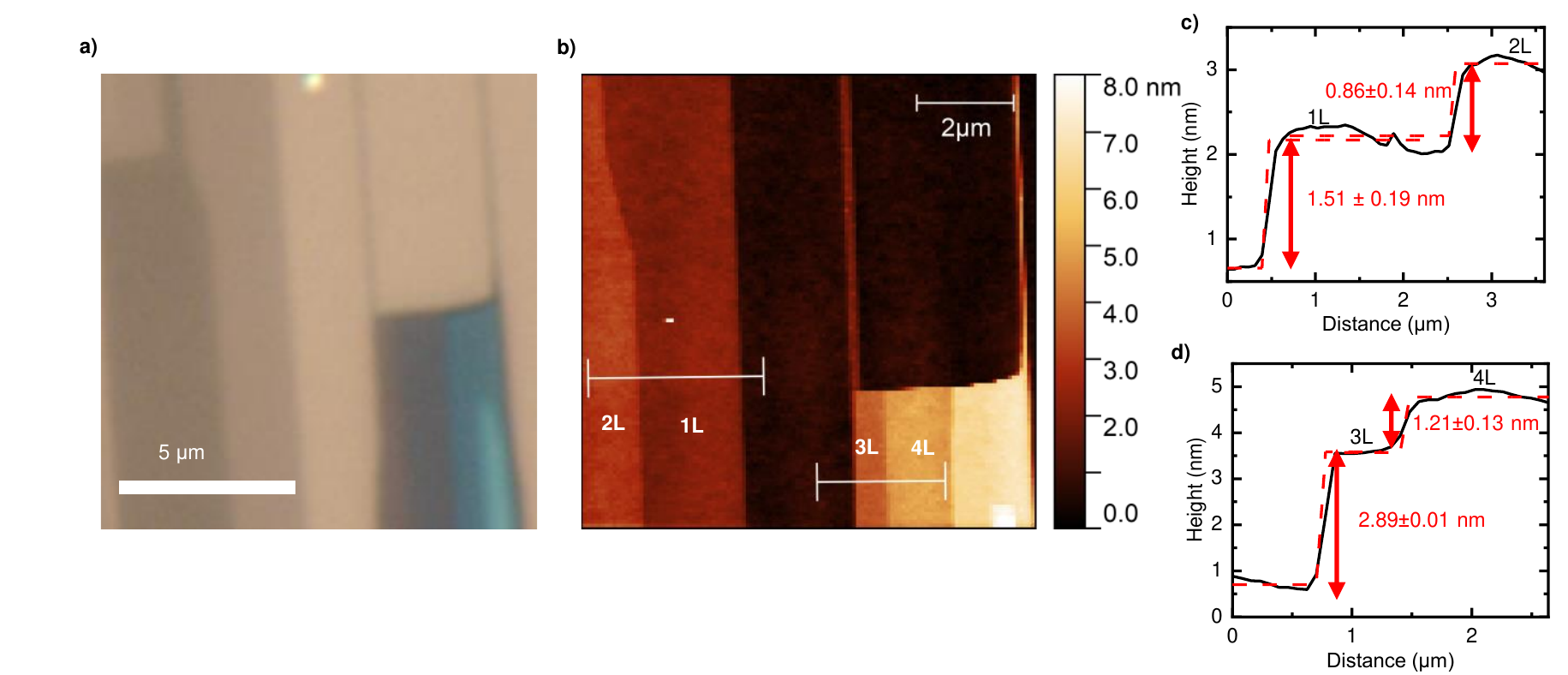}}
\renewcommand{\figurename}{SI Fig.|}
\caption{\label{SIAFM}
\textbf{Atomic force microscopy} 
\textbf{a}, Optical microscope image of 
1L-4L CrSBr flake and 
\textbf{b}, corresponding AFM image with \textbf{c} and \textbf{d}, thickness profiles. Here we measure the monolayer thickness as $\SI{1.51}{\nano\meter}$, contrasting from the expected $\SI{0.791}{\nano\meter}$ thickness for CrSBr~\cite{Lee.2021}. Since flakes are exfoliated in an ambient environment, subsequently after cleaning with isopropanol, it is likely that organic residues and water molecules are confined between the CrSBr/SiO${_2}$ interface. This would result in an increase in measured thickness. A similar effect is commonly observed for other substrate supported 2D material interfaces, such as MoS${_2}$/SiO${_2}$.~\cite{Palleschi2020} 
}
\end{figure*}
%##############################################################################
%

%
%###################### Figure ###############################################
\begin{figure*}[ht]
\scalebox{\figurescale}{\includegraphics[width=1\linewidth]{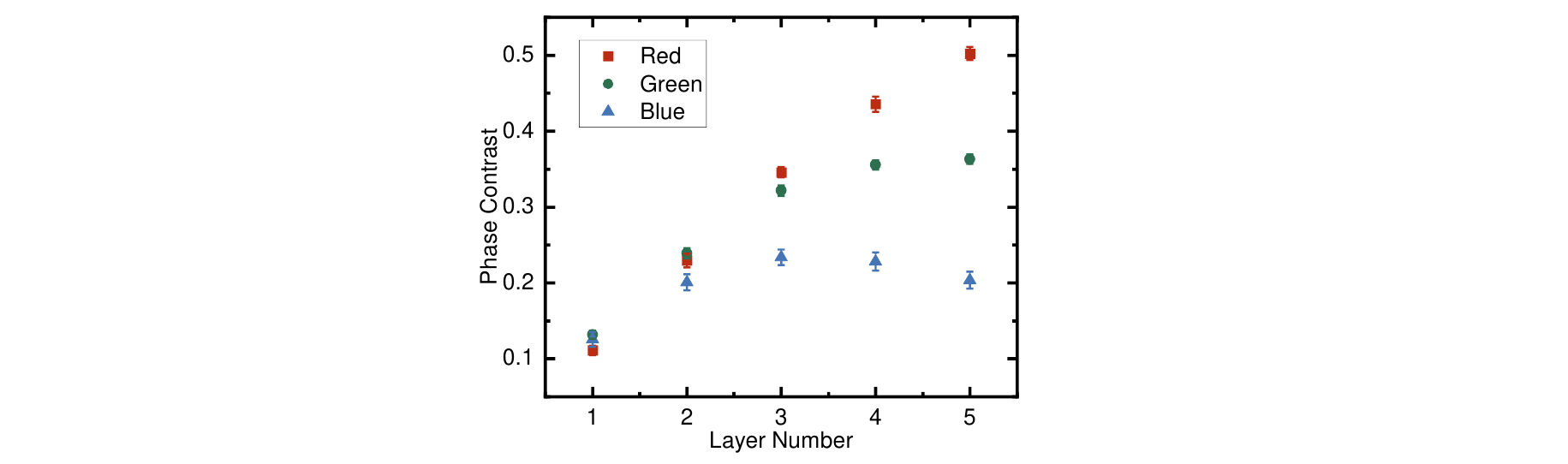}}
\renewcommand{\figurename}{SI Fig.|}
\caption{\label{SIContrast}
\textbf{Optical phase contrast.} Red, green, and blue optical phase contrast ($C = (R_{on}-R_{off})/(R_{on}+R_{off})$) obtained from optical microscope images as a function of CrSBr flake layer number. Because of the clear linear relationship and equidistant steps of the red channel and the layer number, we use the red channel for flake thickness identification for 1L to 5L. 
}

\end{figure*}
%##############################################################################
%

\newpage

\section{Calculated phonon modes.}

%
%###################### Figure ###############################################
\begin{figure*}[ht]
\scalebox{\figurescale}{\includegraphics[width=1\linewidth]{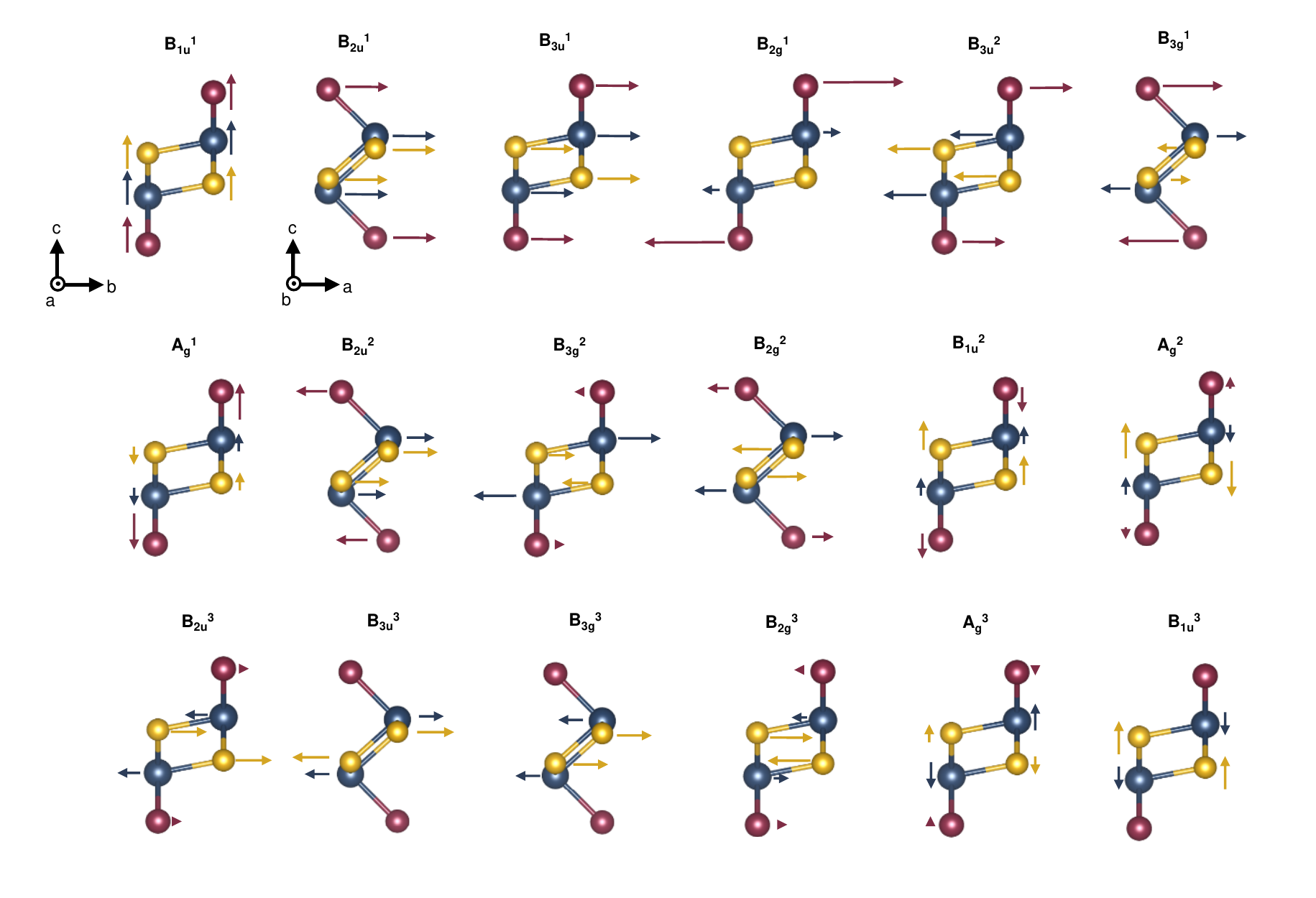}}
\renewcommand{\figurename}{SI Fig.|}
\caption{\label{PhononModes}
\textbf{Phonon modes in monolayer CrSBr.}, Schematic of the atomic displacements corresponding to the 18 phonon modes for monolayer CrSBr obtained from ab initio calculations. 
}

\end{figure*}
%##############################################################################
%

\newpage

\section{Calculated layer-dependent Raman spectra.}

%
%###################### Figure ###############################################
\begin{figure*}[ht]
\scalebox{\figurescale}{\includegraphics[width=1\linewidth]{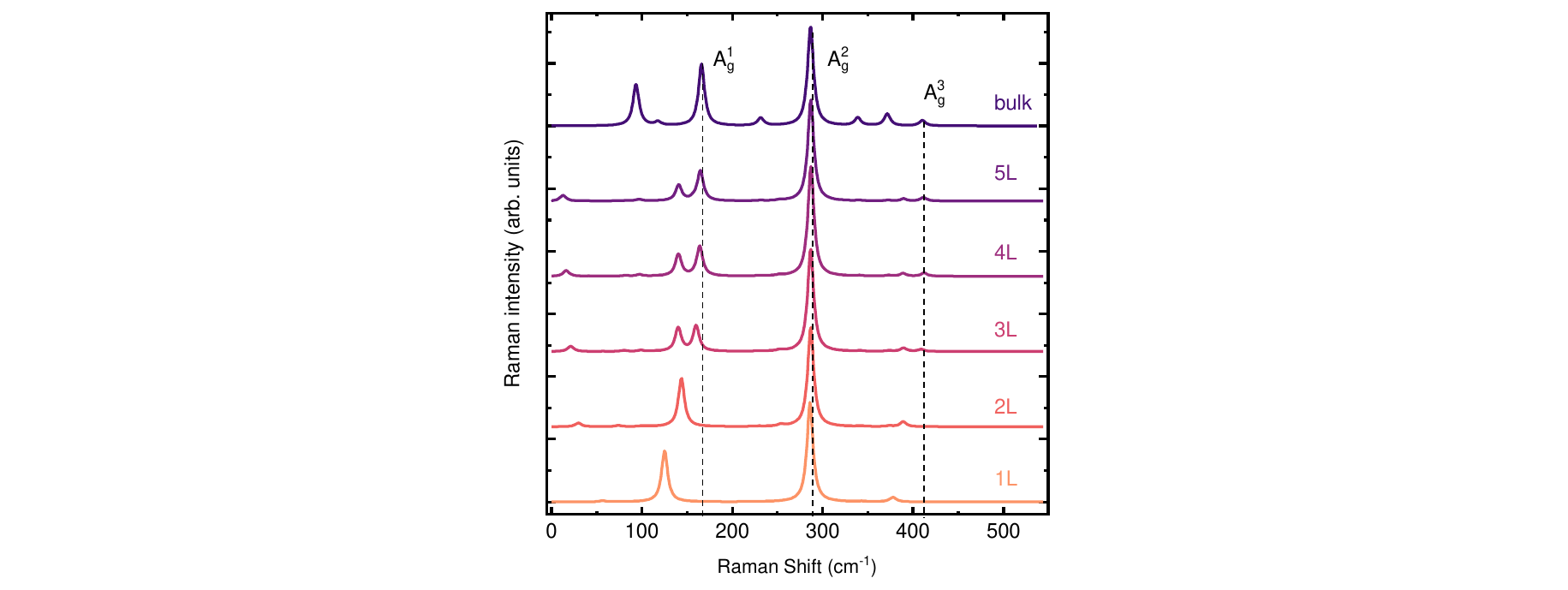}}
\renewcommand{\figurename}{SI Fig.|}
\caption{\label{PhononModes}
\textbf{Calculated Raman spectra of 1L and few-layer CrSBr.} 1L to bulk ab initio calculated Raman spectra. The A$_g^1$, A$_g^2$ and A$_g^3$ mode positions for the bulk are highlighted. The A$_g^1$ shows a blue-shift for the 1L to the bulk in excellent agreement to the experimental spectra shown in the main text. The splitting of the mode for the 3L, 4L and 5L originates from the surrounding vacuum used as a boundary condition for the calculation. This in turn breaks the symmetry of the outer layers resulting in a second mode at a slightly smaller energy. This directly shows the strong sensitivity of this mode to breaking of symmetry, e.g. by the bromine surface defects that result in the $D1$ mode. In contrast, mode A$_g^2$ and A$_g^3$ show only negligible shift with the layer number owing to the insensitivity from the absence of bromine vibrations and low interlayer vibrational coupling.
}

\end{figure*}
%##############################################################################
%

\newpage

\section{Laser excitation energy dependence and resonance condition.}

%
%###################### Figure ###############################################
\begin{figure*}[ht]
\scalebox{\figurescale}{\includegraphics[width=1\linewidth]{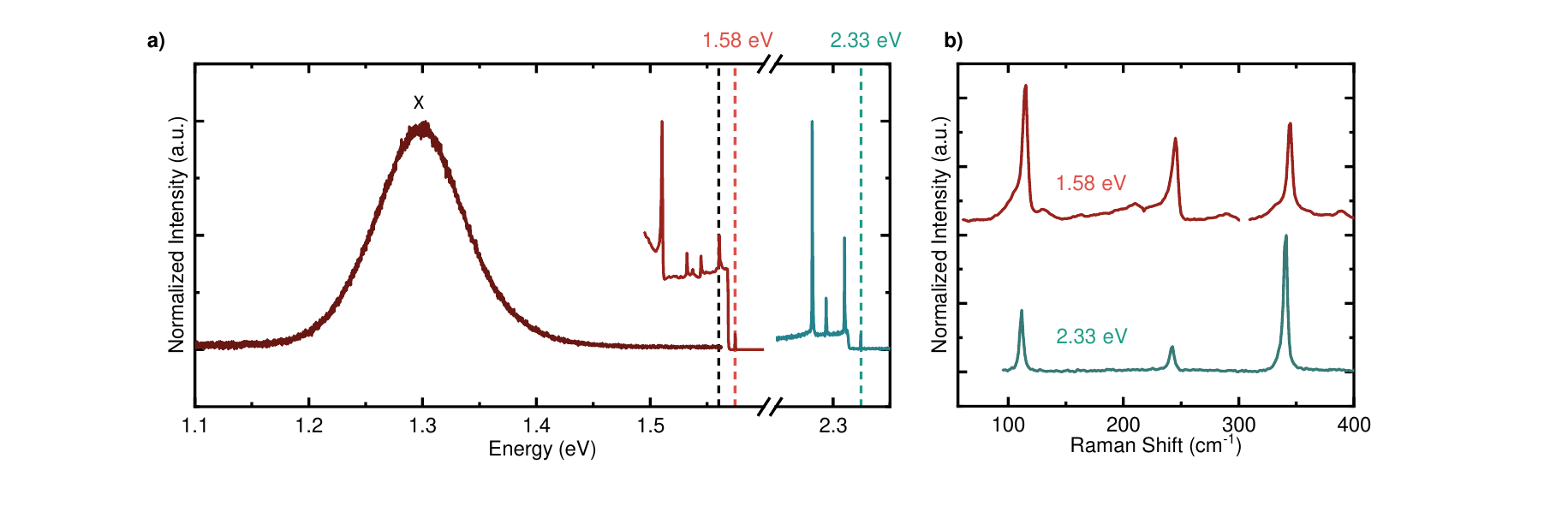}}
\renewcommand{\figurename}{SI Fig.|}
\caption{\label{LaserEnergy}
\textbf{Raman excitation energy dependence.} \textbf{a}, Photoluminescence and Raman spectrum of bulk CrSBr obtained for excitation at $E_L = \SI{1.58}{\electronvolt}$ (red) which is on-resonance with the expected bulk single-particle band gap at E$_{g} \sim \SI{1.56}{\electronvolt}$~\cite{Klein.2022a} represented with the black line, exciton labeled as X, and Raman spectrum obtained at the off-resonant $E_L = \SI{2.33}{\electronvolt}$ excitation plotted as a function of energy prior to any background removal.\textbf{b}, Bulk Raman spectra at on-resonant and off-resonant excitations.}
\end{figure*}
%##############################################################################
%

\newpage

\section{Polarization of resonant Raman modes.}

%
%###################### Figure ###############################################
\begin{figure*}[ht]
\scalebox{\figurescale}{\includegraphics[width=1\linewidth]{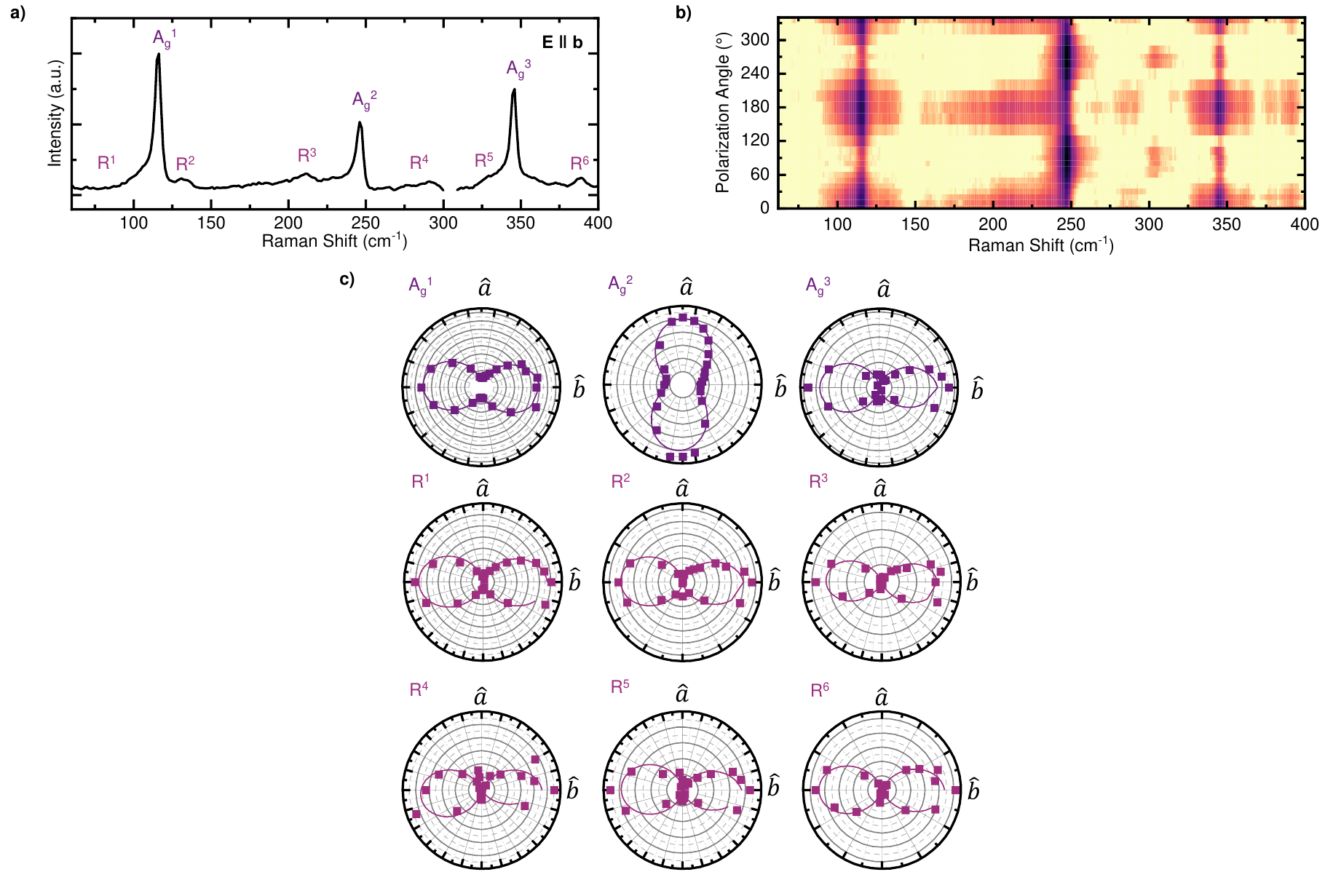}}
\renewcommand{\figurename}{SI Fig.|}
\caption{\label{SIPolar}
\textbf{Polarization of resonant Raman modes in bulk CrSBr.}
\textbf{a}, Bulk Raman spectra aligned along the b axis, \textbf{b}, False color contour plot of angle resolved polarized bulk CrSBr. At $\sim \SI{300}{\per\centi\meter}$ there is an artifact in the spectra related to the SiO$_{2}$/Si substrate \textbf{c} corresponding polar plots of the three A$_{g}$ modes and the eight resonant Raman modes ($R1$-$R8$) displaying a strong polarization response. All data were collected in a linearly co-polarized configuration under resonant excitation conditions ($E_L = \SI{1.58}{\electronvolt}$). 
}

\end{figure*}
%##############################################################################
%

\newpage

\section{Raman spectrum background subtraction.}
%
%###################### Figure ###############################################
\begin{figure*}[ht]
\scalebox{\figurescale}{\includegraphics[width=1\linewidth]{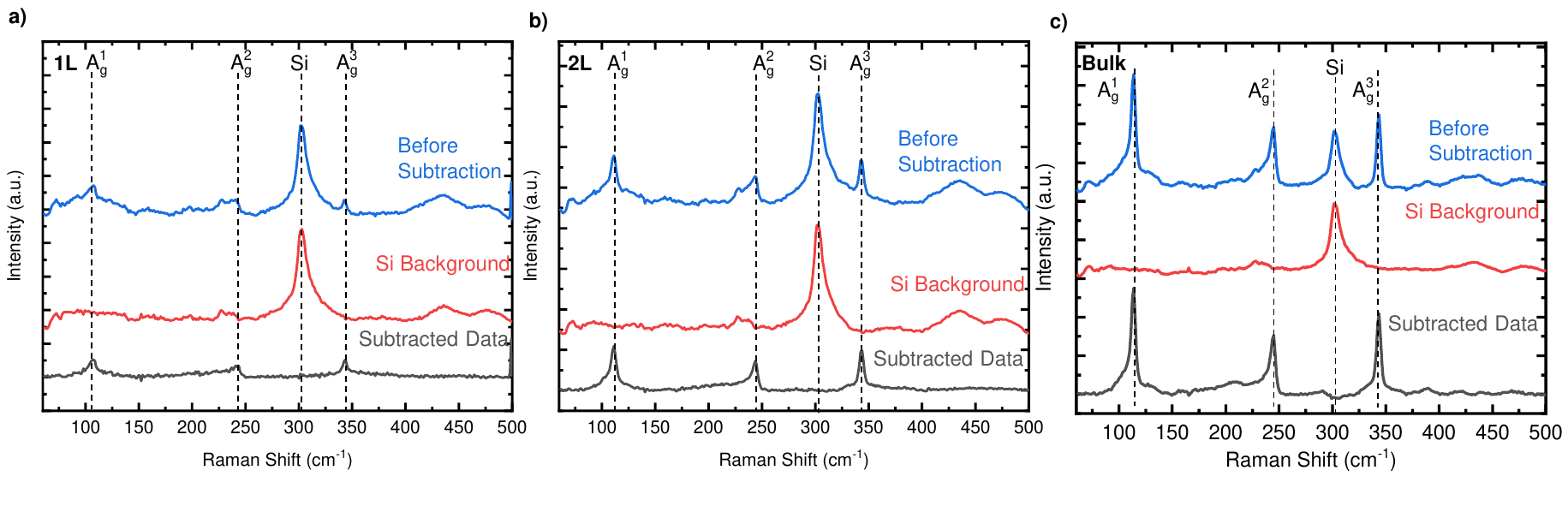}}
\renewcommand{\figurename}{SI Fig.|}
\caption{\label{SISiSubtract}
\textbf{Raman background subtraction.}
Raman spectra before background subtraction, Raman of SiO$_{2}$/Si substrate background and spectra after background subtraction for \textbf{a} 1L, \textbf{b} 2L, and \textbf{c} bulk CrSBr with $E_L = \SI{1.58}{\electronvolt}$.As seen most prominently in \textbf{c}, substrate subtraction may yield an artifact in the spectra at $\sim \SI{300}{\per\centi\meter}$. In such cases, we exclude the artifact points from the plotted spectra for clarity. 
}
\end{figure*}
%##############################################################################
%

\newpage

\section{Time evolution and stability.}

%
%###################### Figure ###############################################
\begin{figure*}[ht]
\scalebox{\figurescale}{\includegraphics[width=1\linewidth]{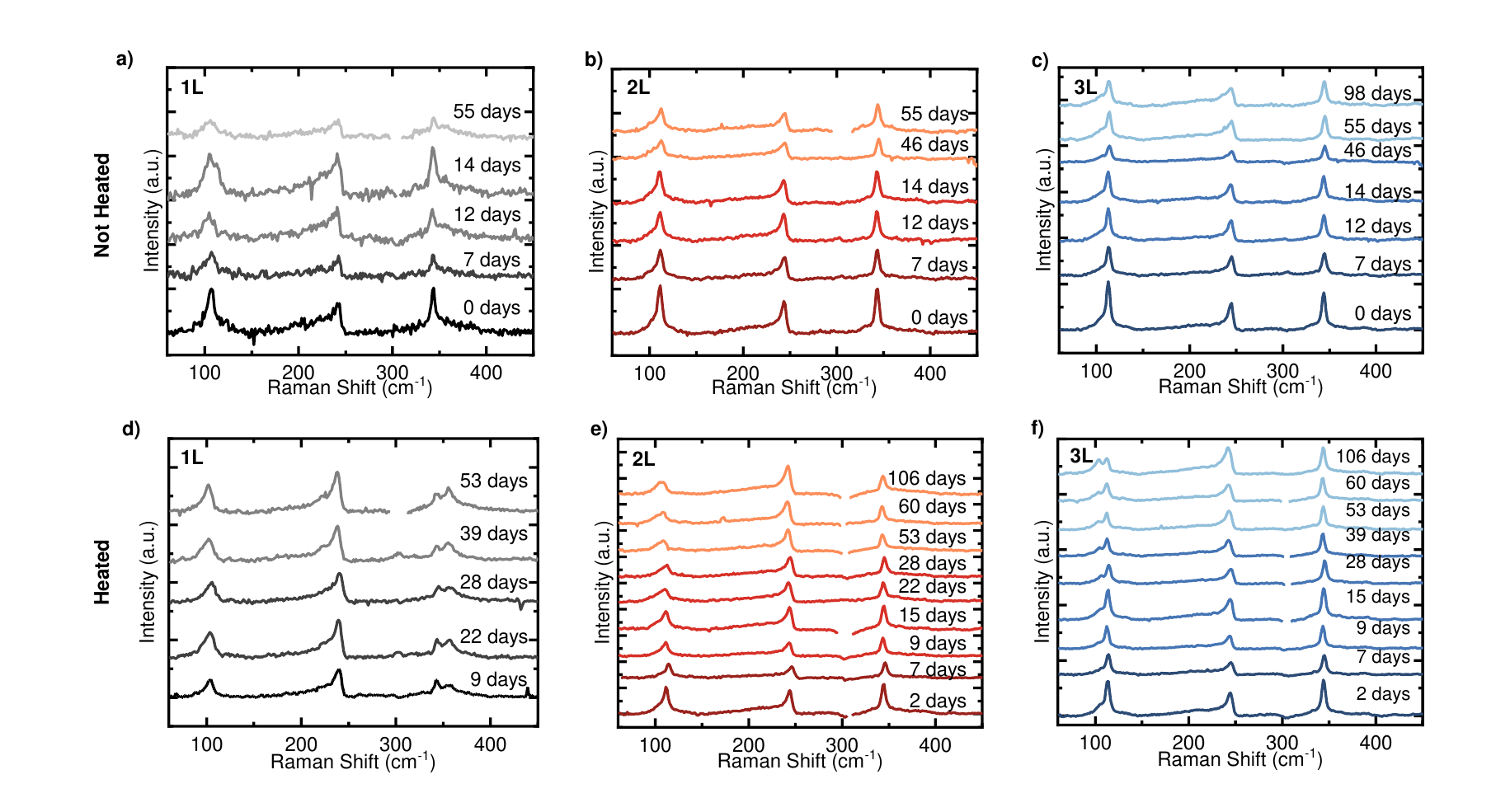}}
\renewcommand{\figurename}{SI Fig.|}
\caption{\label{SIAirStable}
\textbf{Air stability of CrSBr.}
\textbf{a}, 1L \textbf{b}, 2L and \textbf{c}, 3L flakes over several days months of continuous air exposure for flakes as exfoliated and \textbf{d-f} after heating to $\SI{105}{\degree}$ for 2 minutes. Overall, we observe that heated flakes exhibit more degradation over comparable air exposure times. As heating may induce more bromine vacancies, heated flakes are likely more reactive to air and water. 
}
\end{figure*}
%##############################################################################
%

\newpage

%
%###################### Figure ###############################################
\begin{figure*}[ht]
\scalebox{\figurescale}{\includegraphics[width=1\linewidth]{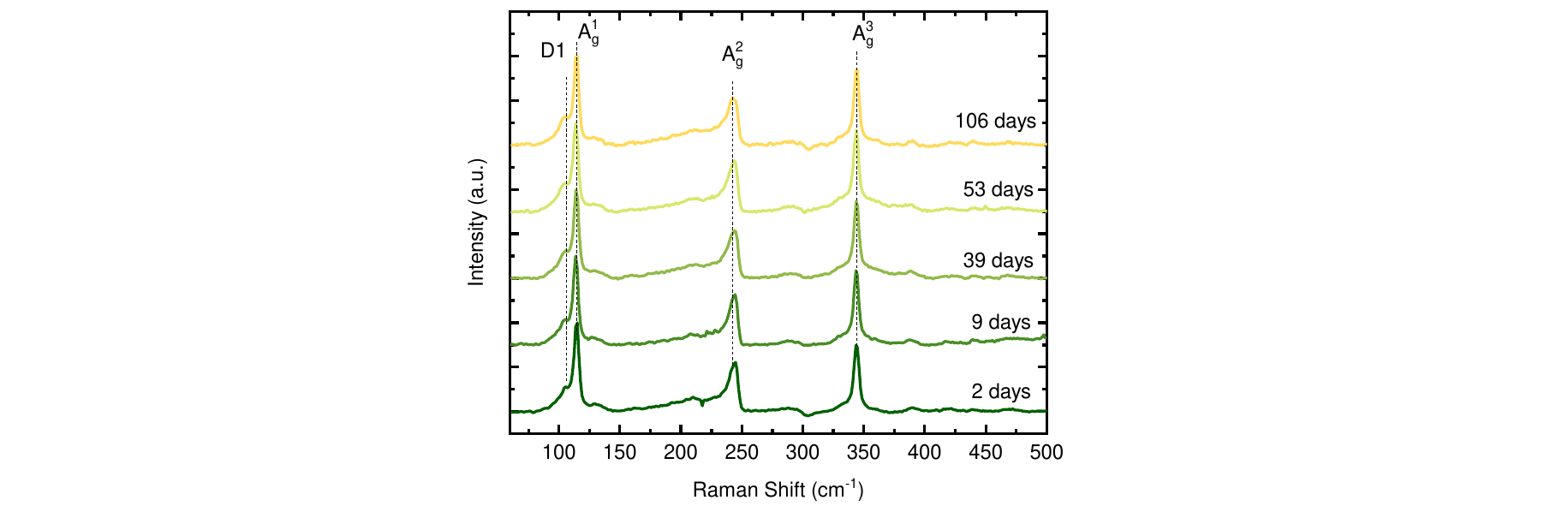}}
\renewcommand{\figurename}{SI Fig.|}
\caption{\label{SI5L}
\textbf{Defect mode $D1$ in 5L CrSBr}
Raman spectra of a 5L CrSBr flake left in air over a period of several days. Similar to flakes in Figure 2, the $D1$ defect mode becomes more prominent over time left in air.
}
\end{figure*}
%##############################################################################

\newpage

%
%###################### Figure ###############################################
\begin{figure*}[ht]
\scalebox{\figurescale}{\includegraphics[width=1\linewidth]{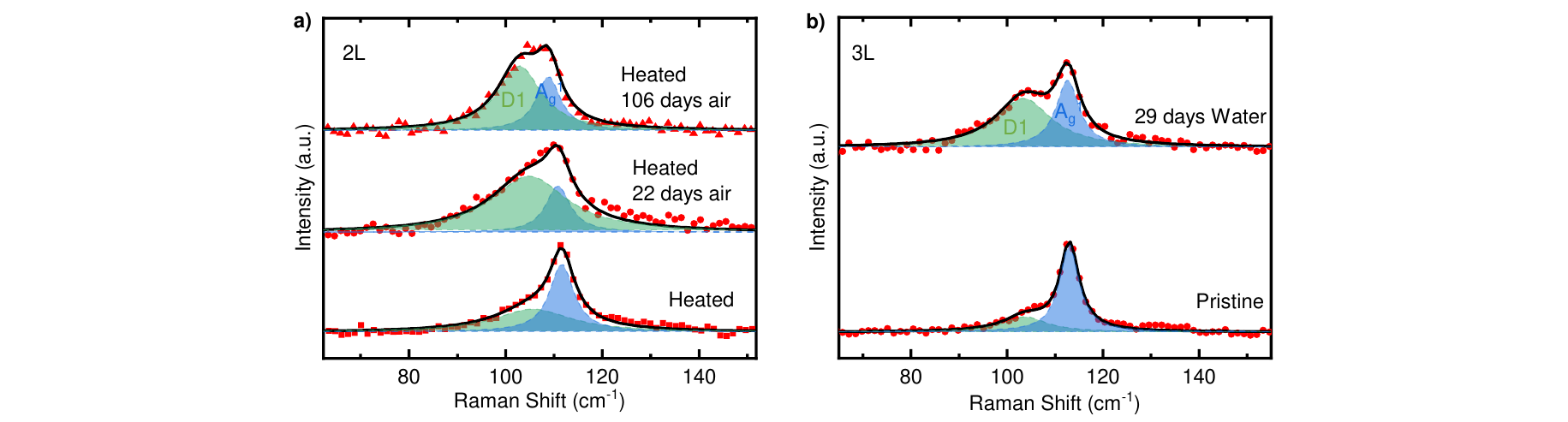}}
\renewcommand{\figurename}{SI Fig.|}
\caption{\label{SI2L3LD1}
\textbf{Defect mode $D1$ in 2L and 3L CrSBr.}
\textbf{a}, Defect $D1$ and A$_{g}^{1}$ mode with Lorentzian fits for various defective flakes. For the 2L, the frequency difference between the $D1$ and A$_{g}^{1}$ mode is only $\sim 6 cm^{-1}$ yielding a more overlapping, asymmetric line shape. This contrasts with the 3L, which has a $D1$ and A$_{g}^{1}$ frequency difference of $\sim 9 cm^{-1}$ resulting in a more distinct mode splitting.
}
\end{figure*}
%##############################################################################

\newpage

%
%###################### Figure ###############################################
\begin{figure*}[ht]
\scalebox{\figurescale}{\includegraphics[width=1\linewidth]{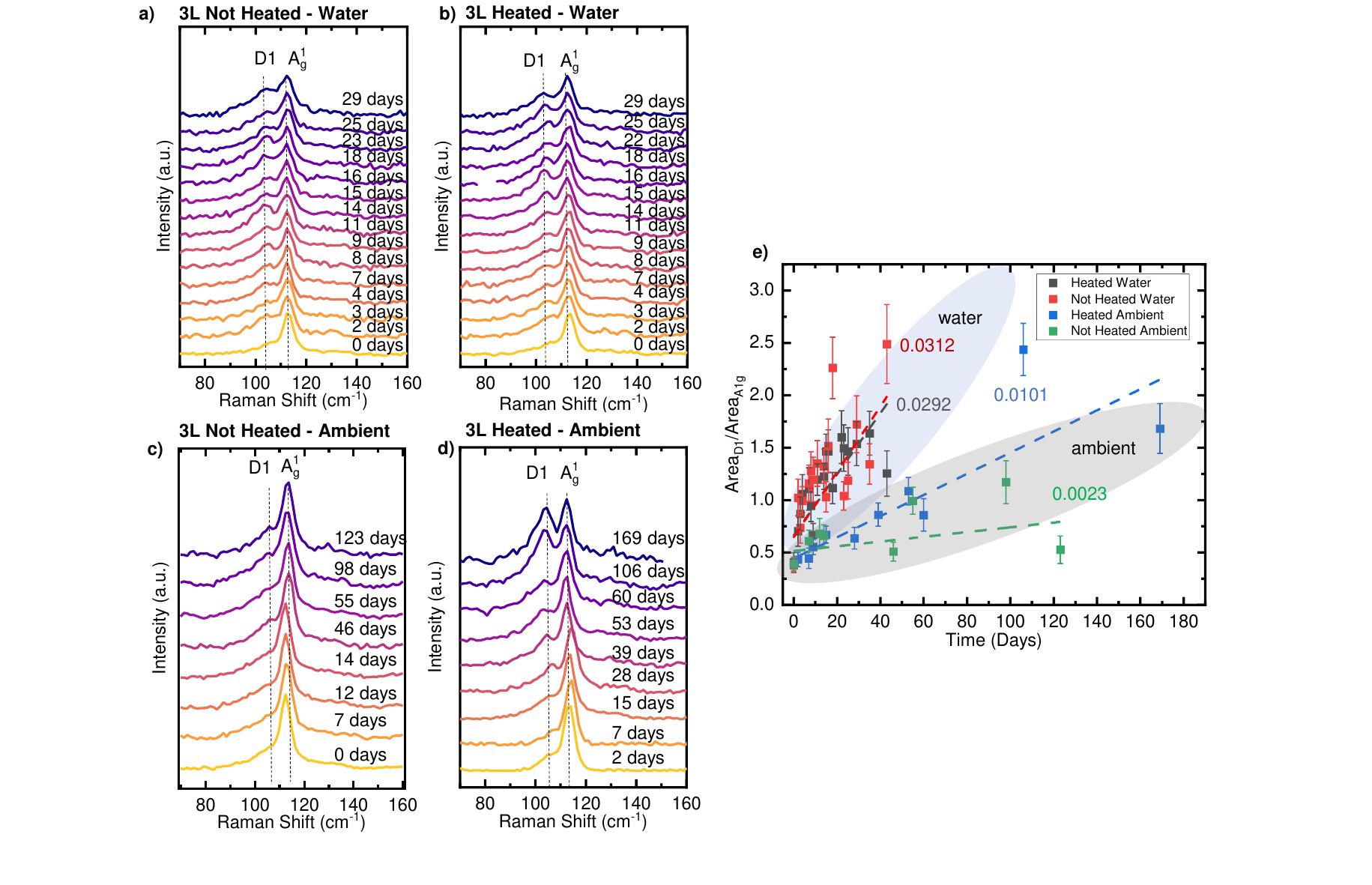}}
\renewcommand{\figurename}{SI Fig.|}
\caption{\label{SI3LRatio}
\textbf{Development of surface related defect mode $D1$ in a 3L CrSBr.}
\textbf{a}, Raman spectra of CrSBr 3L over a period of 29 days submerged in water as exfoliated and \textbf{b}, heated to 75 °C for 5 minutes after exfoliation. \textbf{c} Raman spectra of CrSBr 3L over a period of several days left in ambient as exfoliated and \textbf{d} heated to 105 °C for 2 minutes.  \textbf{e}, the area of the $D1$ mode divided over the area of the A$_{g}^{1}$ mode as a function of time submerged in water (corresponding to flakes in \textbf{a}, and \textbf{b}) or left in air (corresponding to flakes in \textbf{c}, and \textbf{d})  For each flake, there is a positive linear relationship between the area ratio as a function of time, with the slopes listed on the plot. The slope is significantly higher for flakes submerged in water versus those left in air, indicating that submerging in water accelerates the degradation process and defect formation.
}
\end{figure*}
%##############################################################################

\newpage

\section{Raman spectra of helium ion irradiated 4L and bulk.}

%
%###################### Figure ###############################################
\begin{figure*}[ht]
\scalebox{\figurescale}{\includegraphics[width=1\linewidth]{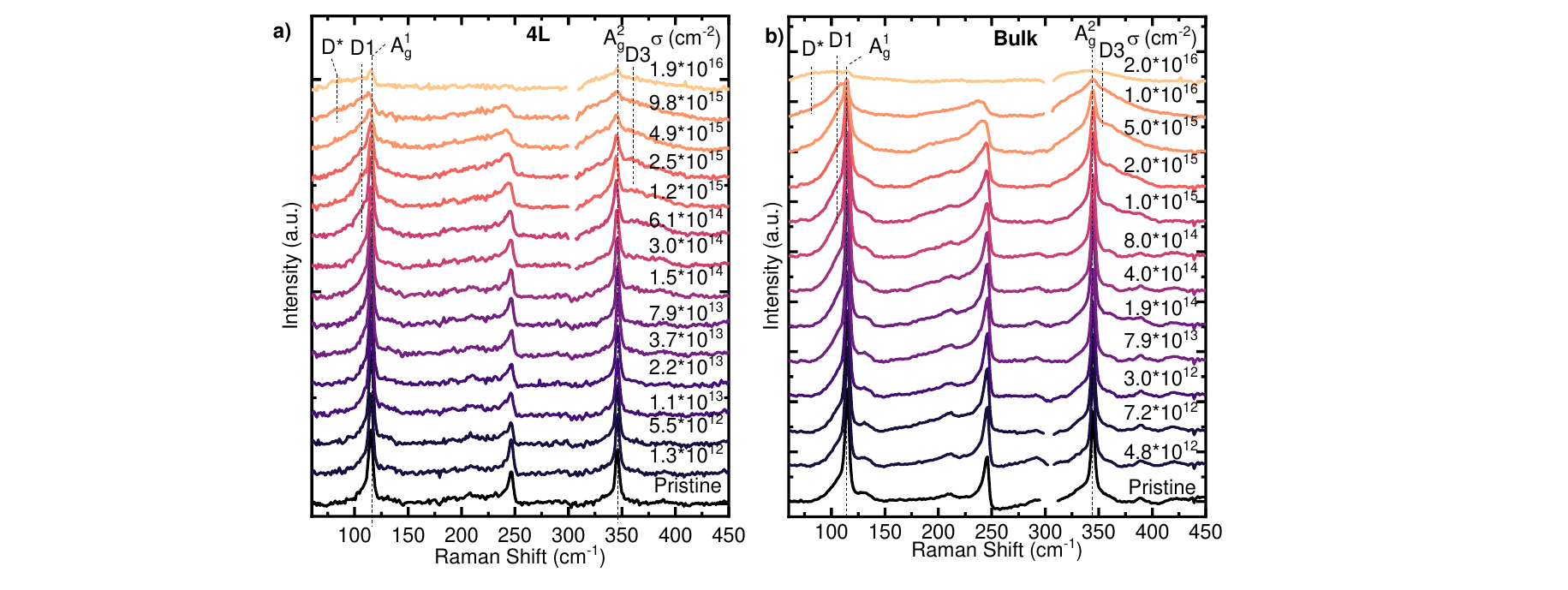}}
\renewcommand{\figurename}{SI Fig.|}
\caption{\label{SIHeliumIon}
\textbf{Helium ion irradiated 4L and bulk CrSBr.}
Waterfall plot of Raman spectra of helium ion irradiated \textbf{a}, 4L and \textbf{b} bulk ($\sim$ 10 layers thick) CrSBr flakes.  
}
\end{figure*}

\newpage

\section{Nano Auger electron spectroscopy.}

%
%###################### Figure ###############################################
\begin{figure*}[ht]
\scalebox{\figurescale}{\includegraphics[width=1\linewidth]{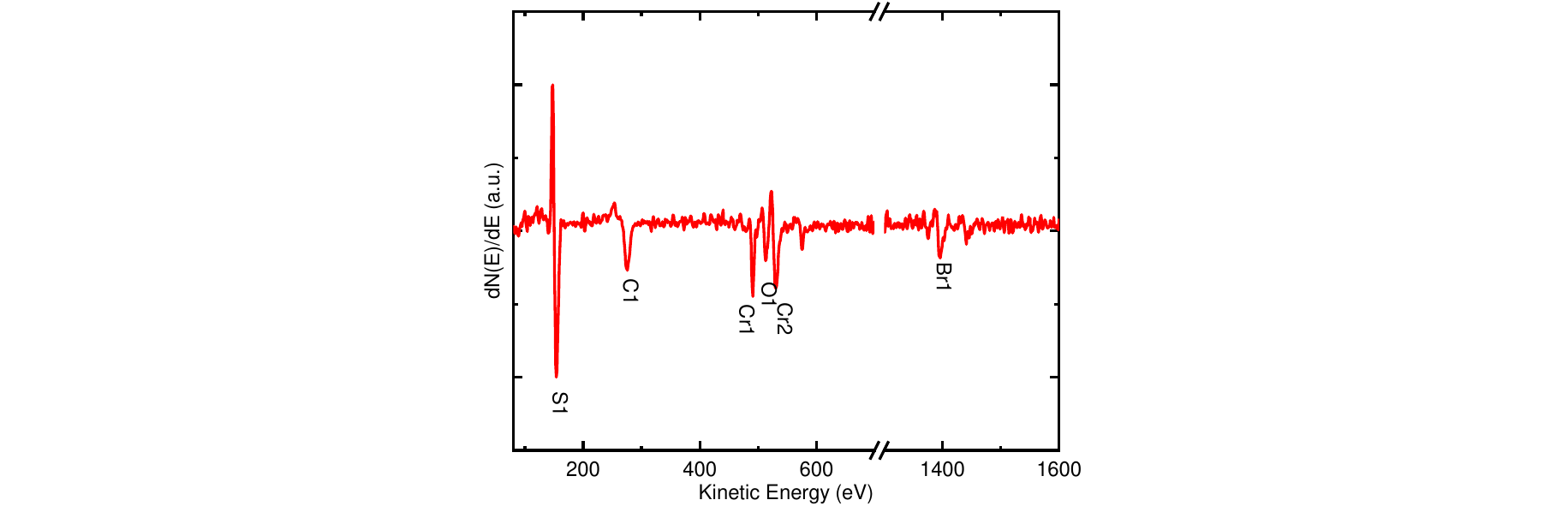}}
\renewcommand{\figurename}{SI Fig.|}
\caption{\label{SIAugerSpectra}
\textbf{Nano Auger electron spectroscopy of bulk CrSBr.}
Nano Auger spectrum with labeled characteristic elemental peaks for a pristine CrSBr bulk flake.
}
\end{figure*}
%##############################################################################
%

\begin{figure*}[ht]
\scalebox{\figurescale}{\includegraphics[width=1\linewidth]{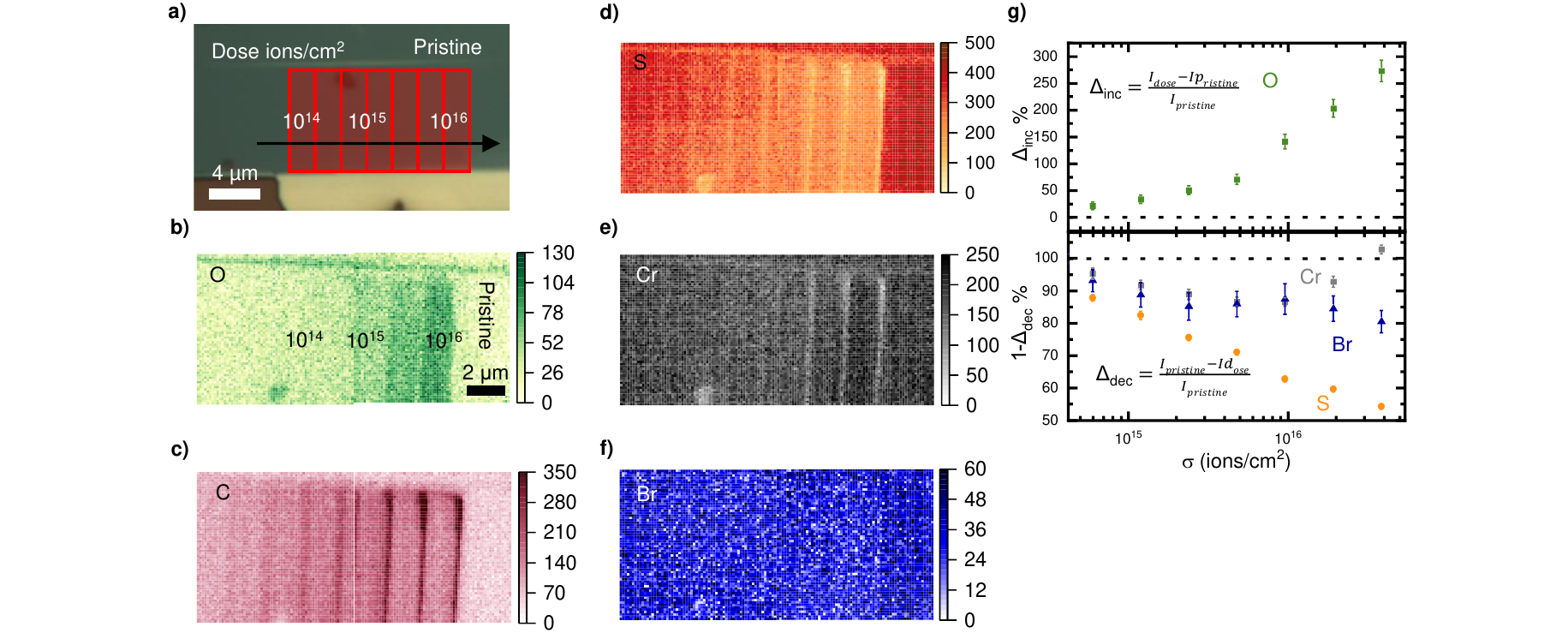}}
\renewcommand{\figurename}{SI Fig.|}
\caption{\label{SIAugerMaps}
\textbf{Irradiated Nano Auger elemental maps of bulk CrSBr.}
\textbf{a}, Optical microscope image of a bulk CrSBr flake and highlighted dose fields irradiated with different helium ion doses.
\textbf{b}, Elemental Maps obtained via Nano Auger electron spectroscopy for oxygen, \textbf{c} carbon, \textbf{d} sulfur,  \textbf{e} chromium and  \textbf{f} bromine. \textbf{g}, Percent increase of oxygen and carbon and percent decrease of sulfur, chromium, and bromine as a function of helium ion dose. 
}
\end{figure*}

Given the high spatial resolution of the Nano Auger microscope, are able to obtain elemental maps down to the $\sim \SI{100}{\nano\meter}$ range for a CrSBr flake irradiated at dose fields ranging from 10$^{14}$ to 10$^{15}$ ions/$\SI{}{\per\centi\meter\squared}$.Nano Auger electron spectroscopy is also a highly surface sensitive technique with a high depth resolution of $\sim \SI{3}{\nano\meter}$.   Figure~\ref{SIAugerMaps} displays the elemental maps collected for each of the five elements of interest, with the percent increase $\Delta_{inc}$ or percent decrease $\Delta_{dec}$ of each element as a function of dose. The percent increase or decrease is shown below

\begin{equation}
\label{eq:inc}
    \Delta_{inc} = \frac{I_{dose}-I_{pristine}}{I_{pristine}}
\end{equation}

\begin{equation}
\label{eq:dec}
    \Delta_{dec} = - \Delta_{inc}
\end{equation}

with I$_{pristine}$ and I$_{dose}$ defined as the average characteristic peak intensity over the pristine region and specified dose field, respectively. 

At each pixel in the maps, the intensity of the characteristic peak corresponding to the specified element is represented. We observe a clear weakening in the sulfur Auger peak signal, thereby decrease in sulfur concentration at the surface of the flake, with increasing helium ion dose. This suggests that helium ion irradiation creates sulfur vacancies. Chromium has a strong Auger peak, however we observe little change in chromium composition even at high helium ion doses. This implies that chromium is not fully removed from the CrSBr during helium ion irradiation. It is possible that helium ion irradiation displaces the chromium atoms into interstitial positions or generally disturbs the crystalline structure. On the other hand, chromium atoms are more stable in the lattice as suggested by a high defect formation energy.~\cite{Klein.2022a} This is generally promising given that of the net magnetic moment arises from contributions and interactions from chromium atoms. This would suggest that CrSBr still could have stable magnetic ordering even at high defect doses. We also do not observe a clear trend in bromine concentration with helium ion dose. As the characteristic bromine peak has a rather weak Auger signal, changes in bromine concentration are likely difficult to detect. Additionally, the intrinsic bromine vacancy concentration in CrSBr is already so high~\cite{Klein.2022a} that irradiation may not change the amount of bromine vacancies at the surface drastically. 

%
%###################### Figure ###############################################
%##############################################################################
%

Given that oxygen and carbon are abundant in ambient conditions and can be detected using Auger, elemental maps of oxygen and carbon are obtained. Auger measurements are carried out under high vacuum, shortly after helium ion irradiation. However, irradiated surfaces are still exposed to ambient conditions for a few minutes between transfer from the helium ion microscope to the Nano Auger chamber. Auger maps show a clear increase in oxygen and carbon concentration with helium ion dose. It is noted that the increase in carbon concentration may be due to carbon deposition during helium ion irradiation. The increase in oxygen concentration at higher doses suggests that the defective surfaces are ideal host sites for oxygen atoms. The chemisorption of atomic or molecular species onto CrSBr vacancy sites is a viable defect mechanism. For instance, oxygen has been shown to highly effectively passivate sulfur vacancies in MoS$_{2}$ and WS$_2$ at ambient conditions.~\cite{Pet2018,Barja2019,Schuler.2019} Given that we observe the simultaneous increase in oxygen and decrease in sulfur, it is likely that that the oxygen-sulfur substitution is a defect process in CrSBr. 

\newpage

\section{Low- and room temperature Raman spectra of 2L and 3L.}

%
%###################### Figure ###############################################
\begin{figure*}[ht]
\scalebox{\figurescale}{\includegraphics[width=1\linewidth]{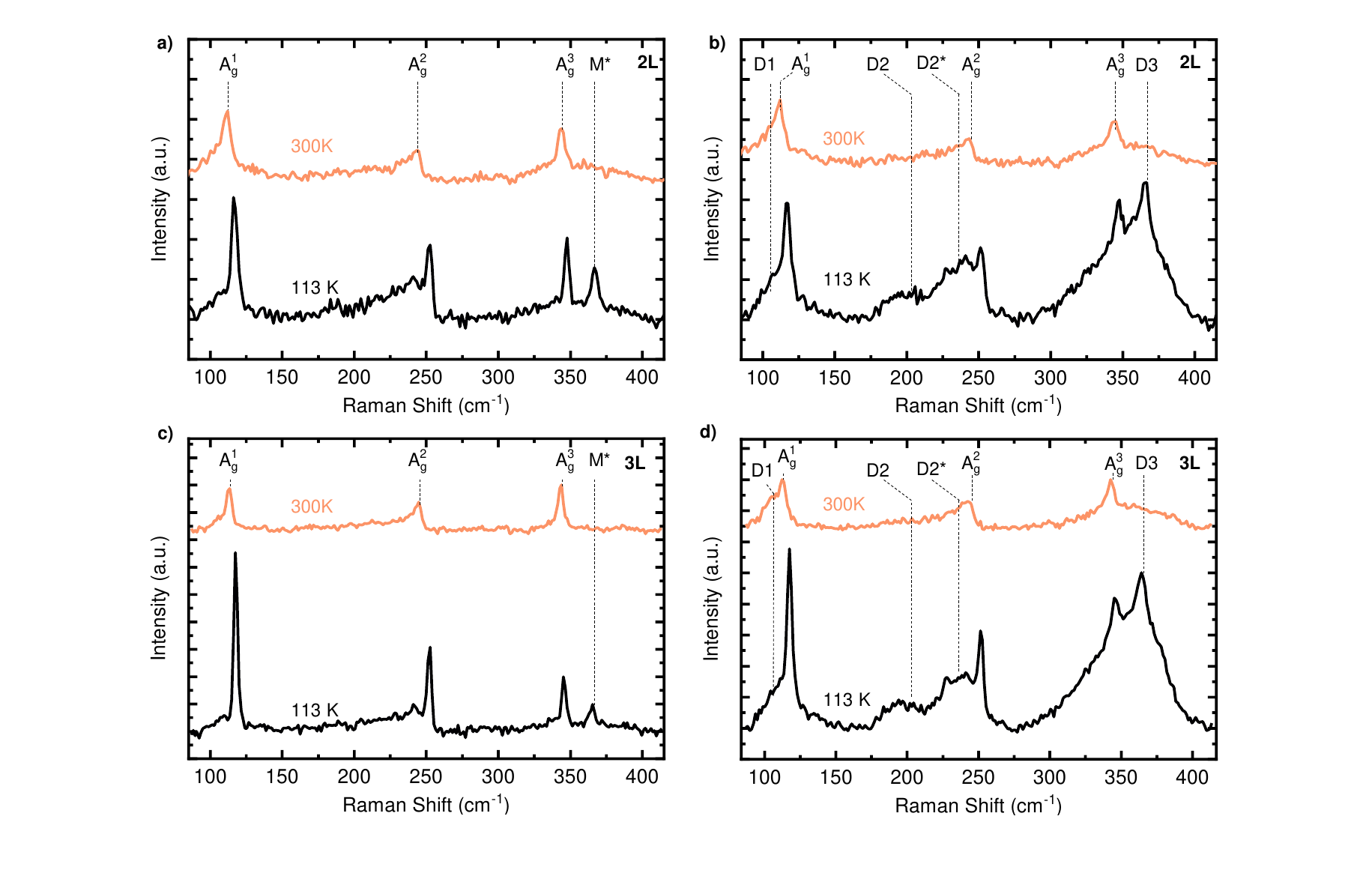}}
\renewcommand{\figurename}{SI Fig.|}
\caption{\label{2L low temperature}
\textbf{Pronounced spin-phonon coupling in pristine and He$^+$ irradiated 2L CrSBr.}
\textbf{a} Resonant Raman spectra collected at $E_L = \SI{1.58}{\electronvolt}$ of pristine 2L CrSBr at $\SI{113}{\kelvin}$ and $\SI{300}{\kelvin}$ and \textbf{b} 2L He$^+$ irradiated CrSBr irradiated at a dose of $\sigma = 8 \cdot 10^{14} \SI{}{\per\centi\meter\squared}$ at $\SI{113}{\kelvin}$ and $\SI{300}{\kelvin}$ \textbf{c} pristine and \textbf{d} irradiated 3L CrSBr flakes at the same temperature and dose for comparison
}
\end{figure*}

%##############################################################################

%
%##############################################################################
%               Acknowledgements & Contributions
%##############################################################################
%
%###############################################################################
%								Additional information
%###############################################################################
%
%###############################################################################
%								BIBLIOGRAPHY
%##############################################################################
%
%\FloatBarrier
\bibliographystyle{apsrev}
\bibliography{full}% Produces the bibliography via BibTeX.